\documentclass[twocolumn,floatfix,prl]{revtex4}%
\usepackage{graphicx}%
\usepackage{amsmath,amssymb}%
\usepackage{amsfonts}
\usepackage{amssymb}
\usepackage{mathrsfs}
\usepackage{color}
\usepackage{bm}
\def\d{{\partial}}
\def\s{{\sigma}}
\def\e{{\epsilon}}
\def\k{{ {\bm k} }}

\def\q{{ {\bm q} }}

\def\B{{ {\bm B} }}

\def\0{{ {\bm 0} }}
\def\w{{\omega}}
\def\a{{\alpha}}
\def\b{{\beta}}

\def\A{{ {\rm A} }}
\def\B{{ {\rm B} }}
\def\C{{ {\rm C} }}
\allowdisplaybreaks[4]

\begin{document}
\title{Drastic magnetic-field-induced chiral current order  
and emergent current-bond-field interplay
in kagome metal AV$_3$Sb$_5$ (A=Cs,Rb,K)}
\author{
Rina Tazai$^1$, Youichi Yamakawa$^2$, and Hiroshi Kontani$^2$
\footnote[0]{$^*$These authors made equal contribution to the work.}
}
\date{\today}

\begin{abstract}
In kagome metals, the chiral current order parameter ${\bm\eta}$
with time-reversal-symmetry-breaking
is the source of various exotic electronic states, 
while the method of controlling the current order
and its interplay with the star-of-David bond order ${\bm\phi}$
are still unsolved.
Here, we reveal that tiny uniform orbital magnetization 
$M_{\rm orb}[{\bm\eta},{\bm\phi}]$ is induced by the chiral current order,
and its magnitude is prominently enlarged 
under the presence of the bond order.
Importantly, 
we derive the magnetic-field ($h_z$)-induced Ginzburg-Landau (GL)
free energy expression
$\Delta F[h_z,{\bm\eta},{\bm\phi}]\propto-h_z M_{\rm orb}[{\bm\eta},{\bm\phi}]$,
which enables us to elucidate the field-induced
current-bond phase transitions in kagome metals.
The emergent current-bond-$h_z$ trilinear coupling term in the free energy,
$-3m_1 h_z{\bm\eta}\cdot{\bm\phi}$,
naturally explains the characteristic magnetic field sensitive 
electronic states in kagome metals,
such as the field-induced current order
and the strong interplay between the bond and current orders.
The GL coefficients of $\Delta F[h_z,{\bm\eta},{\bm\phi}]$ derived from the realistic multiorbital model are appropriate to explain various experiments.
Furthermore, we present a natural explanation for the 
drastic strain-induced increment of the current order transition
temperature $T_{\rm c}$ reported by a recent experiment.
\end{abstract}

\address{
$^1$Yukawa Institute for Theoretical Physics, Kyoto University, 
Kyoto 606-8502, Japan \\
$^2$ Department of Physics, Nagoya University,
Furo-cho, Nagoya 464-8602, Japan. 
}

\sloppy

\maketitle

\section*{Introduction}
Recent discovery of unconventional quantum phases in metals
has led to a new trend of condensed matter physics.
Exotic charge-density-wave orders
and unconventional superconductivity
in geometrically frustrated kagome metal AV$_3$Sb$_5$ (A=Cs,Rb,K)
have been attracting increasing attention
\cite{kagome-exp1,kagome-exp2}.
The 2$\times$2 (inverse) star-of-David order,
which is presumably the triple-$\q$ ($3Q$) bond order (BO),
occurs at $T_{\rm BO}\approx 100$ K at ambient pressure
\cite{STM1,STM2}.
The BO is the time-reversal-symmetry (TRS) preserving
modulation in the hopping integral,
$\delta t_{ij}^{\rm b}={\rm real}$.
\cite{Thomale2013,SMFRG,Thomale2021,Neupert2021,Balents2021,Nandkishore,Tazai-kagome}.
Below $T_{\rm BO}$, nodeless superconductivity occurs for A=Cs
\cite{Roppongi,SC2},
which is naturally explained based on the BO fluctuation mechanism
proposed in Ref. \cite{Tazai-kagome}.

In kagome metals, unusual TRS breaking (TRSB) phase
without long-range spin orders
has been reported by $\mu$-SR \cite{muSR3-Cs,muSR2-K,muSR4-Cs,muSR5-Rb},
Kerr rotation \cite{birefringence-kagome,Yonezawa},
field-tuned chiral transport \cite{eMChA} measurements
and STM studies under magnetic field \cite{STM1,eMChA}.
The transition temperature $T_{\rm TRSB}$ is still under debate.
Although it is close to $T_{\rm BO}$ in many experiments,
the TRSB order parameter is strongly magnified at 
$T^* \approx 35$K for A=Cs
\cite{eMChA,muSR3-Cs,muSR4-Cs} and
$T^* \approx 50$K for A=Rb
\cite{muSR5-Rb}.
Recently, magnetic torque measurement reveals the nematic order
with TRSB at $T^{**}\approx130$K \cite{Asaba},
while no TRSB was reported by recent Kerr rotation study
\cite{Kapitulnik}.
A natural candidate is the correlation driven
TRSB hopping integral modulation:
$\delta t_{ij}^{\rm c} ={\rm imaginary}$.
This order accompanies topological charge-current
 \cite{Haldane}
that gives the giant anomalous Hall effect (AHE) 
\cite{AHE1,AHE2}.

Theoretically, the BO and the current order emerge
in the presence of sizable off-site electron correlations
in Fe-based and cuprate superconductors
\cite{Onari-SCVC,Yamakawa-FeSe,Onari-TBG,Tsuchiizu1,Tsuchiizu4,Chubukov-PRX2016,Fernandes-rev2018,Kontani-AdvPhys,Davis-rev2013}
and in kagome metals
\cite{Thomale2021,Neupert2021,Balents2021,Tazai-kagome,Tazai-kagome2,PDW-kagome,Fernandes-GL,Thomale-GL,Nat-g-ology}.
\textcolor{black}{
Notably, strong off-site interaction 
(due to the off-site Coulomb repulsion or the BO fluctuations)
gives rise to the charge current order 
\cite{Neupert2021,Balents2021,Nandkishore,Tazai-kagome2}.
Based on the GL free-energy analysis, interesting
bond+current nematic ($C_2$) coexisting phases have been discussed
in two-dimensional (2D) and three-dimensional (3D) models 
\cite{Balents2021,Tazai-kagome2,Fernandes-GL,Thomale-GL,Nat-g-ology}.
Experimentally, the nematic state is actually observed by 
the elastoresistance
\cite{elastoresistance-kagome},
the scanning birefringence
\cite{birefringence-kagome},
and the STM 
\cite{STM2}
measurements.
}

In kagome metals, outer magnetic field $h_z$
drastically modifies the electronic states.
The chirality of the charge-current is aligned
under very tiny $|h_z|\sim1$ Tesla
according to the measurements of AHE \cite{AHE1,AHE2}
and field-tuned chiral transport \cite{eMChA}.
In addition, the amplitude of the loop current 
is strongly magnified by applying small $h_z \ (\gtrsim1{\rm Tesla})$
\cite{muSR4-Cs,muSR2-K,muSR5-Rb}.
Very recent transport measurement of highly symmetric fabricated 
CsV$_3$Sb$_5$ micro sample 
\cite{Moll-hz}
reveals that current-order state is drastically enlarged by the small $h_z$.
These drastic $h_z$-dependences are the hallmarks of the TRSB state,
and it is important to understand the coupling between 
the current order, chirality and the magnetic field in kagome metals.

\begin{figure}[htb]
\includegraphics[width=.90\linewidth]{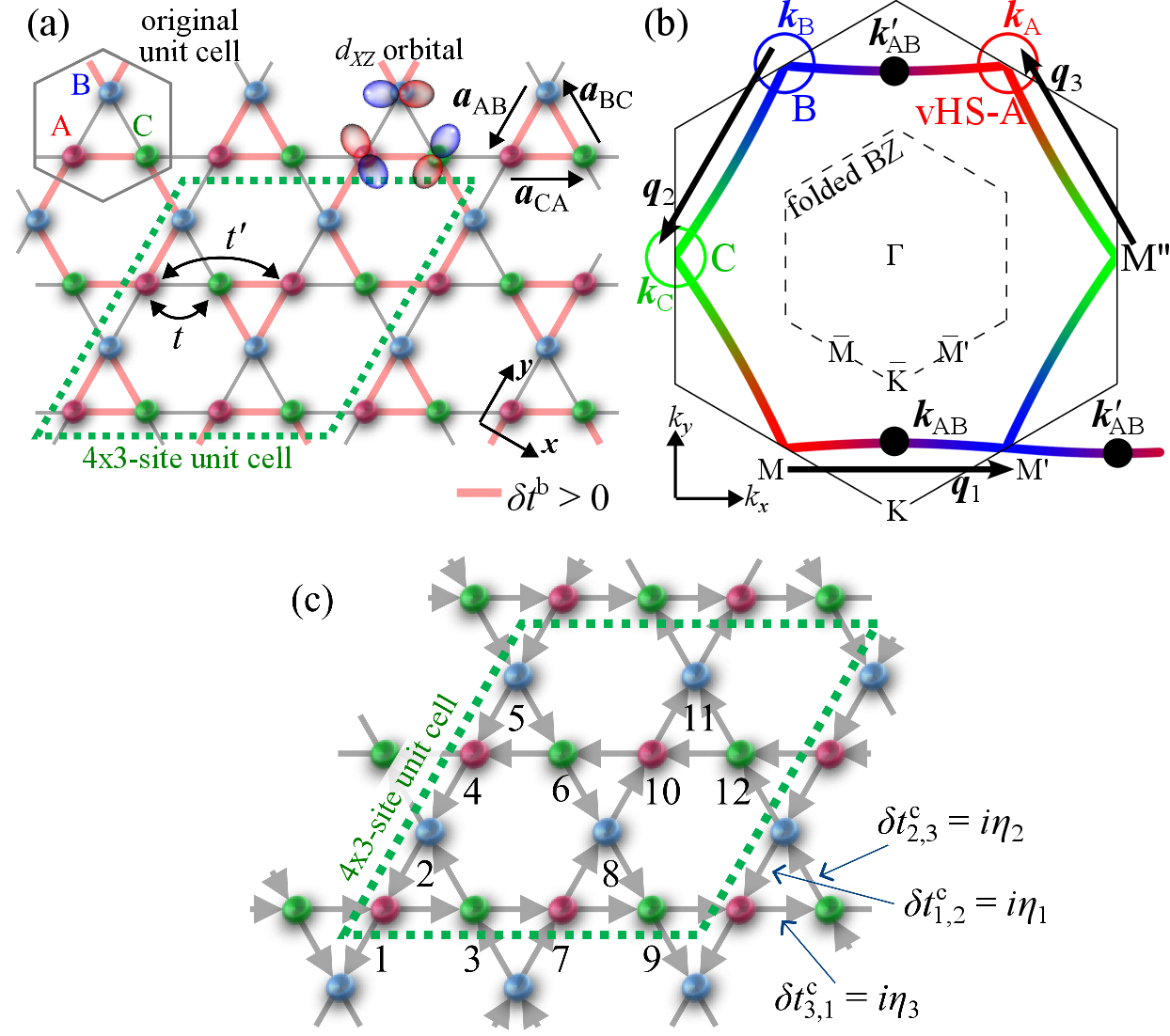}
\caption{
(a) Kagome lattice tight-binding model.
The $3Q$ BO
${\bm \phi}=(\phi,\phi,\phi)/\sqrt{3}$ $(\phi>0)$ is depicted.
The $d_{XZ}$-orbital on each sublattice is drawn.
(b) FS at the vHS filling $n_{\rm vHS}$.
Three vHS points $\k_{\rm A}$, $\k_{\rm B}$ 
and $\k_{\rm C}$ are respectively composed of 
$A$ (red), $B$ (blue), and $C$ (green) orbitals.
The inter-vHS nesting vectors $\q_n$ ($n=1-3$) are shown.
(c) $3Q$ current order 
${\bm \eta}=(\eta_1,\eta_2,\eta_3)$ is depicted.
Note that $(\eta_1,\eta_2,\eta_3)$ and $(\eta_1,-\eta_2,-\eta_3)$
are the same bulk states.
}
\label{fig:fig1}
\end{figure}
In this paper, 
we reveal that the $3Q$ loop-current order parameters
accompany tiny orbital magnetization, $M_{\rm orb}$,
and its magnitude is drastically enlarged 
under the presence of the bond order.
Importantly, 
we derive the $h_z$-induced Ginzburg-Landau (GL) free energy 
expression $\Delta F$,
which is useful to study nontrivial phase transitions 
under the magnetic field.
The emergent current-bond-$h_z$ trilinear coupling term in $\Delta F$
not only explains the origin of novel field-induced 
chiral symmetry breaking 
\cite{muSR4-Cs,muSR2-K,muSR5-Rb}
but also provides useful hints to control the charge current 
in kagome metals.
In addition, the ``strain-induced increment of $T_{\rm TRSB}$ 
reported in Ref. \cite{Moll-hz} is naturally understood.

In the present study of $M_{\rm orb}$,
we mainly focus on the current order in the 
$3d_{XZ}$-orbital (in $b_{3g}$-representation) of V ion,
which has been intensively studied previously
\cite{Thomale2021,Neupert2021,Balents2021,Nandkishore,Tazai-kagome2}.
However, other $3d$-orbitals (especially $3d_{YZ}$-, 
$d_{X^2-Y^2}$-, $d_{3Z^2-R^2}$-orbitals) also form the large
Fermi surfaces (FSs) with the van-Hove singularity (vHS) points 
near the Fermi level.
The impact of these non-$b_{3g}$ orbitals on the current order 
has also been studied in Refs. \cite{Fernandes-GL,Nat-g-ology}.
The present field-induced GL theory does not depend on the 
$d$-orbital character of the current order parameter.
We calculate the GL coefficients $m_n$ ($n=1-3$)
for various $d$-orbital current order states
based on the first-principles kagome metal models.
For a fixed order parameter,
$m_n$ is large when the FS reconstruction 
due to the current order parameters occurs near the vHS points.

\section*{Model Hamiltonian with current and bond orders}
Here, we study the kagome-lattice tight-binding model 
with $b_{3g}$ (or $d_{XZ}$) orbitals shown in Fig. \ref{fig:fig1} (a).
Each unit-cell is composed of three sublattices A, B, C.
We set the nearest-neighbor hopping integral $t=-0.5[{\rm eV}]$.
In addition, we introduce the 
nearest intra-sublattice hopping $t'=-0.02[{\rm eV}]$
shown in Fig. \ref{fig:fig1} (a) to avoid the perfect nesting.
(Hereafter, the unit of the energy is eV.)
The FS at the vHS filling,
$n=n_{\rm vHS}=2.55$ per 3-site unit cell, is shown in Fig. \ref{fig:fig1} (b).
Then, the FS lies on the three vHS points 
($\k_{\A}$, $\k_{\B}$, $\k_{\C}$),
each of which is composed of a single sublattice A, B, or C.
This simple three-orbital model 
well captures the main pure-type FS in kagome metals
\cite{STM1,ARPES-VHS,ARPES-band,ARPES-Lifshitz,Sato-ARPES}.

The bond/current order is the modulation
of the hopping integral between $i$ and $j$ atmos
due to the electron correlation, $\delta t_{ij}^{\rm b/c}$.
Theoretically, it is the symmetry breaking in the self-energy,
and it is derived from the density-wave (DW) equation
\cite{Tazai-kagome,Tazai-kagome2}.
The wavevectors of the bond and current orders
correspond to the inter-sublattice nesting vectors $\q_n$ ($n=1-3$) 
in Fig. \ref{fig:fig1} (b)
according to previous theoretical studies
\cite{Thomale2021,Neupert2021,Balents2021,Tazai-kagome,Tazai-kagome2,PDW-kagome}.
The triple $Q$ ($3Q$) current order between the nearest atoms
is depicted in Fig. \ref{fig:fig1} (c).
The form factor (=normalized $\delta t_{ij}^{\rm c}$)
with ${\bm q}={\bm q}_1$, $f_{ij}^{(1)}$, is 
$+i$ for $(i,j)$ belongs to the sites $(l,m)=(1,2),(2,4),(4,5),(5,1)$
in Fig. \ref{fig:fig1} (c), and
$-i$ for $(7,8),(8,10),(10,11),(11,7)$.
Odd parity relation $f_{ij}^{(1)}=-f_{ji}^{(1)}$ holds.
Other form factors with ${\bm q}_2$ and ${\bm q}_3$, 
$f_{ij}^{(2)}$ and $f_{ij}^{(3)}$, are also derived from Fig. \ref{fig:fig1} (c).
Using ${\bm f}_{ij}=(f_{ij}^{(1)},f_{ij}^{(2)},f_{ij}^{(3)})$, 
the current order is 
\begin{eqnarray}
\delta t_{ij}^{\rm c}&=& {\bm \eta}\cdot{\bm f}_{ij} ,
\label{eqn:tc}
\end{eqnarray}
%
where
${\bm \eta}\equiv(\eta_1,\eta_2,\eta_3)$ is the set of 
current order parameters with the wavevector $\q_m$.
Also, the BO is given as
\begin{eqnarray}
\delta t_{ij}^{\rm b}&=& {\bm \phi}\cdot{\bm g}_{ij} ,
\label{eqn:tb}
\end{eqnarray}
where ${\bm \phi}\equiv(\phi_1,\phi_2,\phi_3)$ is the set of BO parameters
with the wavevector $\q_m$,
and $g_{ij}^{(m)}=g_{ji}^{(m)}=\pm1$ is the even-parity form factor
for the BO shown in Fig. \ref{fig:fig1} (a).
For example, $g_{ij}^{(1)}=+1 \ [-1]$ for 
$(i,j)$ belongs to $(1,2),(4,5),(8,10),(11,7)$ [$(2,4),(5,1),(7,8),(10,11)$].

The unit cell under the $3Q$ bond/current order
is magnified by $2\times2$ times.
Thus, we analyze the electronic states with the current order
based on the $4\times3$-site kagome lattice model.
The Hamiltonian with the bond+current order is
$\hat{H} =\sum_{\k,l,m,\sigma} h_{lm}(\k) c^{\dagger}_{\k,l,\sigma} c_{\k,m,\sigma}$,
%
%
where $l,m=1\sim12$ 
and $h_{lm}(\k)\ (=h_{ml} (\k)^*)$ is the Fourier 
transform of the hopping integral  
${\tilde t}_{ij}=t_{ij}+\delta t_{ij}^{\rm b}+\delta t_{ij}^{\rm c}$.
Here, $t_{ij}$ is the hopping integral of the original model,
and $\delta t_{ij}^{\rm b (c)}$ is the hopping integral due to the bond (current) 
order in Eq. [\ref{eqn:tb}] (Eq. [\ref{eqn:tc}]).

\section*{Uniform orbital magnetization}

The TRS is broken 
in the presence of the current order $\delta t_{ij}^{\rm c}$.
In this case, the uniform orbital magnetization $M_{\rm orb}$ may appear
due to the finite Berry curvature as pointed out in 
Ref. \cite{Balents2021}.
$M_{\rm orb}$ per V atom in the unit of Bohr magneton
$\mu_{B}=e\hbar/2m_ec$ ($m_e$= free electron mass)
is given as
\cite{Morb-paper1,Morb-paper2}
\begin{eqnarray}
M_{\rm orb}&=&\frac{\mu_{\rm B}}{E_0N_{\rm uc}N}\sum_{\k,\sigma} m(\k),
\label{eqn:Mch} \\
m(\k)&=&\sum_{\a\ne\b}
{\rm Im}({\bm V}^*_{\b\a\k}\times{\bm V}_{\b\a\k})_z
\nonumber \\
&& 
\!\!\!\!\!\!\!\!\!\!\!\!\!\!\!\!\!\!\!\!\!\!\!\!\!\!\!\!\!\!\!\!
\times \left((\e_{\b\k}-\e_{\a\k})n(\e_{\a\k})
-2T\ln[1+e^{-(\e_{\a\k}-\mu)/T}] \right),
\label{eqn:Mch-integ}
\\
&&\!\!\!\!\!\!\!\!
{\bm V}_{\a\b\k}= {\langle \a\k | {\bm \nabla}_\k h_\k | \b\k \rangle}/{(\e_{\a\k}-\e_{\b\k})} ,
\label{eqn:V}
\end{eqnarray}
where $\e_{\a\k}$ is $\a$-th eigenenergy of $4\times3$-site kagome lattice model
in the folded BZ.
$n(\e)$ is Fermi distribution function,
$N_{\rm uc}=12$ is the site number of $2\times2$ unit cell,
$N$ is the $\k$-mesh number, and $E_0= \hbar^2/a^2 m_e$.
$a$ is the unit length in the numerical study,
and we set $a=|{\bm a}_{\rm AB}|$ ($=0.275$ nm in kagome metals).
$E_0=1.0$ eV when $a=0.275$ nm.

\textcolor{black}{
At zero temperature, Eq. [\ref{eqn:Mch}] is rewritten as
\begin{eqnarray}
M_{\rm orb}^{T=0}&=&\frac{\mu_{\rm B}}{E_0N_{\rm uc}N}\sum_{\k,\a<\b}
{\rm Im}\{V^x_{\b\a\k}V^y_{\a\b\k}-(\a\leftrightarrow\b)\}
\nonumber \\
& &\times (\e_{\a\k}+\e_{\b\k}-2\mu)(n^0(\e_{\a\k})-n^0(\e_{\b\k})) ,
\label{eqn:MchT0}
\end{eqnarray}
where $n^0(\e)$ is the Fermi distribution function at $T=0$.
Considering the factor $(n^0(\e_{\a\k})-n^0(\e_{\b\k}))$ with $\a<\b$,
$M_{\rm orb}$ originates from the ``vertical particle-hole (p-h)
excitation'', from $\e_{\a\k}<0$ to $\e_{\b\k}>0$, 
at the same $\k$ in the folded BZ.
}

\textcolor{black}{
The folded FS ($\eta=\phi=0$) and bandstructure in the 
folded Brillouin zone (BZ) at $n= n_{\rm vHS}$ are shown in 
Figs. S1 (a)-(c) in the SI A \cite{SM}.
Here, all vHS points A, B, C in the original BZ in Fig. \ref{fig:fig1} (b)
move to $\Gamma$ point, and they will hybridize each other 
due to the current and/or BO parameter.
When $n\approx n_{\rm vHS}$,
prominent band hybridization occurs near the $\Gamma$ point, as understood in 
Fig. S1 (a).
Then, the bonding (antibonding) band energy at $\Gamma$ point 
is below (above) the Fermi level, as shown in 
Fig. S1 (b).
Because of the factor ${\rm Im}\{\cdots\}\propto |\e_{\a\k}-\e_{\b\k}|^{-2}$ in 
Eq. [\ref{eqn:MchT0}],
$M_{\rm orb}$ becomes large when $n\approx n_{\rm vHS}$.
}

\begin{figure}[htb]
\includegraphics[width=.99\linewidth]{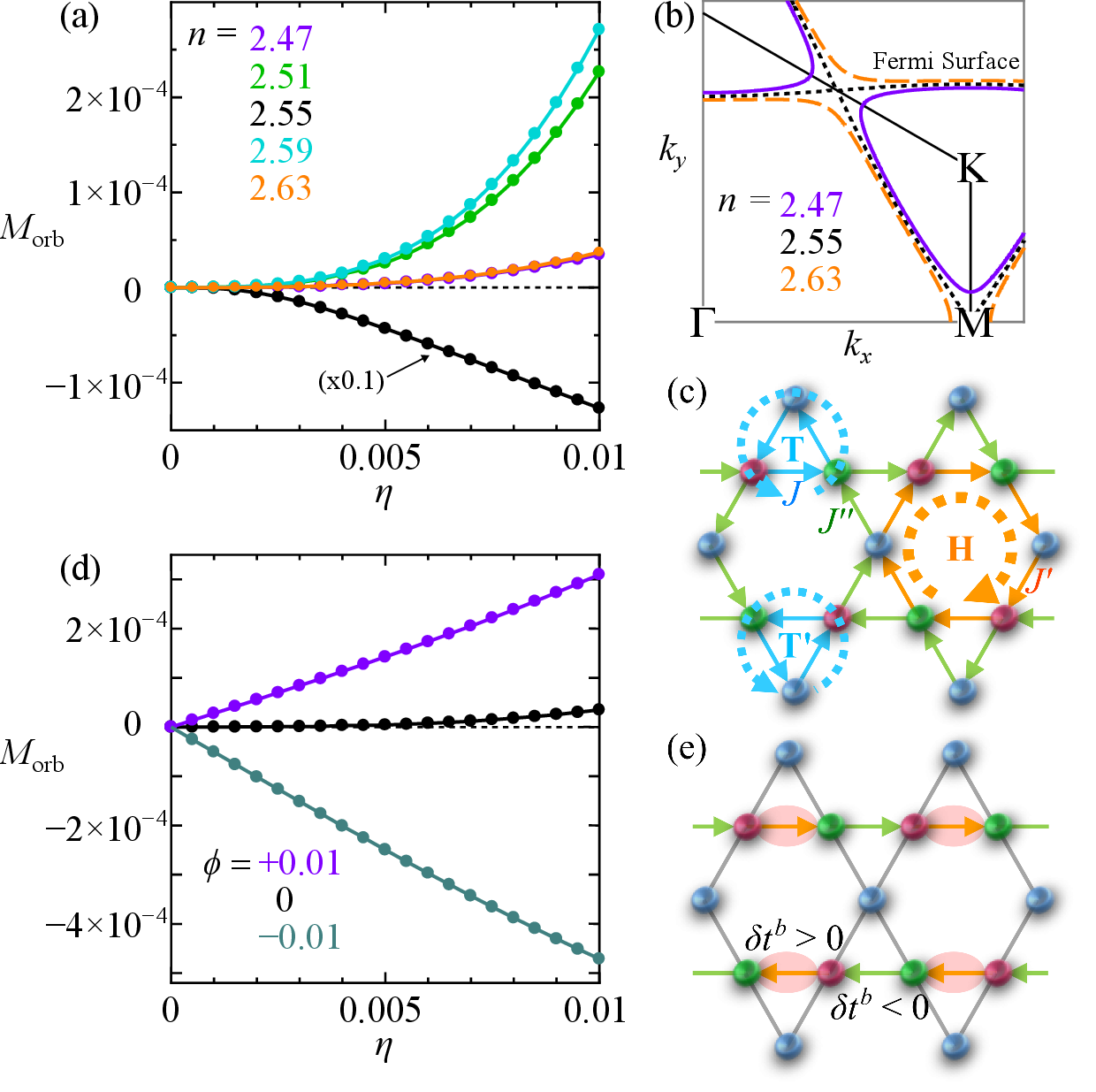}
\caption{
(a) $M_{\rm orb}$ [$\mu_{\rm B}$] per $V$ atom due to the 
$3Q$ current order ${\bm \eta}=(\eta,\eta,\eta)/\sqrt{3}$ 
at $T=1$ meV for $n=2.47\sim2.63$.
$M_{\rm orb} \ (\propto \eta^3)$ is large when $n\approx n_{\rm vHS}$.
(b) FS around $n_{\rm vHS}=2.55$.
(c) Charge-current pattern in real space.
$J\ne J'\ne J''$ in spite of the same order parameter
$\eta_1=\eta_2=\eta_3$.
(d) $M_{\rm orb} \ (\propto \eta)$ due to the coexistence of
$3Q$ current order and $3Q$ BO at $n=2.47$.
(e) $1Q$ bond+current order with finite $M_{\rm orb}$.
Its time-reversal and translation by $2{\bm a}_{\rm AB}$
causes a different state.
}
\label{fig:fig2}
\end{figure}

Here, we calculate $M_{\rm orb}$ [$\mu_{\rm B}$]
near the vHS filling $n_{\rm vHS}=2.55$ at $T=1$ meV,
in the case of $E_0=1.0$ eV.
Figure \ref{fig:fig2} (a) represents the obtained
$M_{\rm orb}$ under the $3Q$ current order
${\bm \eta}=(\eta,\eta,\eta)/\sqrt{3}$ for $n=2.47\sim2.63$.
(The FS without the current order is shown in Fig. \ref{fig:fig2} (b).)
We obtain the relation $M_{\rm orb}\propto \eta^3$,
and it becomes large when the filling is close to $n_{\rm vHS}$.
Therefore, 
the $3Q$ current order state is a weak-ferromagnetic 
(or ferrimagnetic) state.
Because the additional free energy under the magnetic field $h_z$
per 3-site unit cell is 
\begin{eqnarray}
\Delta F=- 3h_z M_{\rm orb} ,
\label{eqn:DeltaF}
\end{eqnarray}
%
the chirality of the current order is aligned under tiny $h_z$.
In other words, the $3Q$ current order is stabilized under $h_z$.
In contrast, $M_{\rm orb}=0$ for $1Q$ and $2Q$ current orders.
$h_z=10^{-4}$ corresponds to $\sim1$ Tesla.



Here, we present a symmetry argument to understand $M_{\rm orb}$
induced by the current order.
Both $1Q$ and $2Q$ current orders are invariant under the 
time-reversal and the translational operations successively.
Therefore, $M_{\rm orb}=0$ due to the  TRS in the bulk state.
In contrast, $M_{\rm orb}$ is nonzero in the $3Q$ current order 
because this state breaks the TRS in the bulk state
\cite{Balents2021}.
To find an intuitive reason,
we calculated the charge-current
along the nearest bonds $J_{ij}$ in the $3Q$ state 
using the method in Refs. \cite{Tazai-cLC,Kontani-AdvPhys},
and found that $|J_{ij}|$ is bond-dependent
in spite of the same order parameter $\eta_1=\eta_2=\eta_3$.
The obtained relation $J\ne J'\ne J''$ 
shown in Fig. \ref{fig:fig2} (c)
indicates that $M_{\rm orb}$ becomes finite
because the magnetic fluxes through triangles and hexagons
do not cancel perfectly.

For general order parameter
${\bm \eta}=(\eta_1,\eta_2,\eta_3)$,
we verified the relation $M_{\rm orb}\propto \eta_1\eta_2\eta_3$
up to the third-order.
In fact, 
based on the perturbation theory with respect to Eq. [\ref{eqn:tc}];
$\eta_m {\hat f}^{(m)}$ ($m=1-3$).
$M_{\rm orb}[{\bm \eta}]$ is expanded as 
$\sum_{pqr} b_{pqr}(\eta_1)^p (\eta_2)^q (\eta_3)^r$
with $p+q+r={\rm odd}$ 
because $M_{\rm orb}[{\bm \eta}]$ is an odd function of ${\bm \eta}$.
\textcolor{black}{
In addition, $b_{pqr}$ can be nonzero only when 
$p\q_1+ q\q_2+r\q_3={\bm 0}$ (modulo original reciprocal vectors),
because we study the nonlinear response of the uniform ($\q={\bm 0}$) magnetization due to the potential $\eta_m f^{(m)}$ with wavevector $\q_m$ in the original unit cell.
(See the SI B for a more detailed explanation \cite{SM}.)
In Fig. \ref{fig:fig2} (a), 
$M_{\rm orb}$ originates from the third-order term $b_{111}$,
which is allowed because of the momentum conservation 
$\q_1+\q_2+\q_3={\bm 0}$.
}



\begin{figure}[htb]
\includegraphics[width=.99\linewidth]{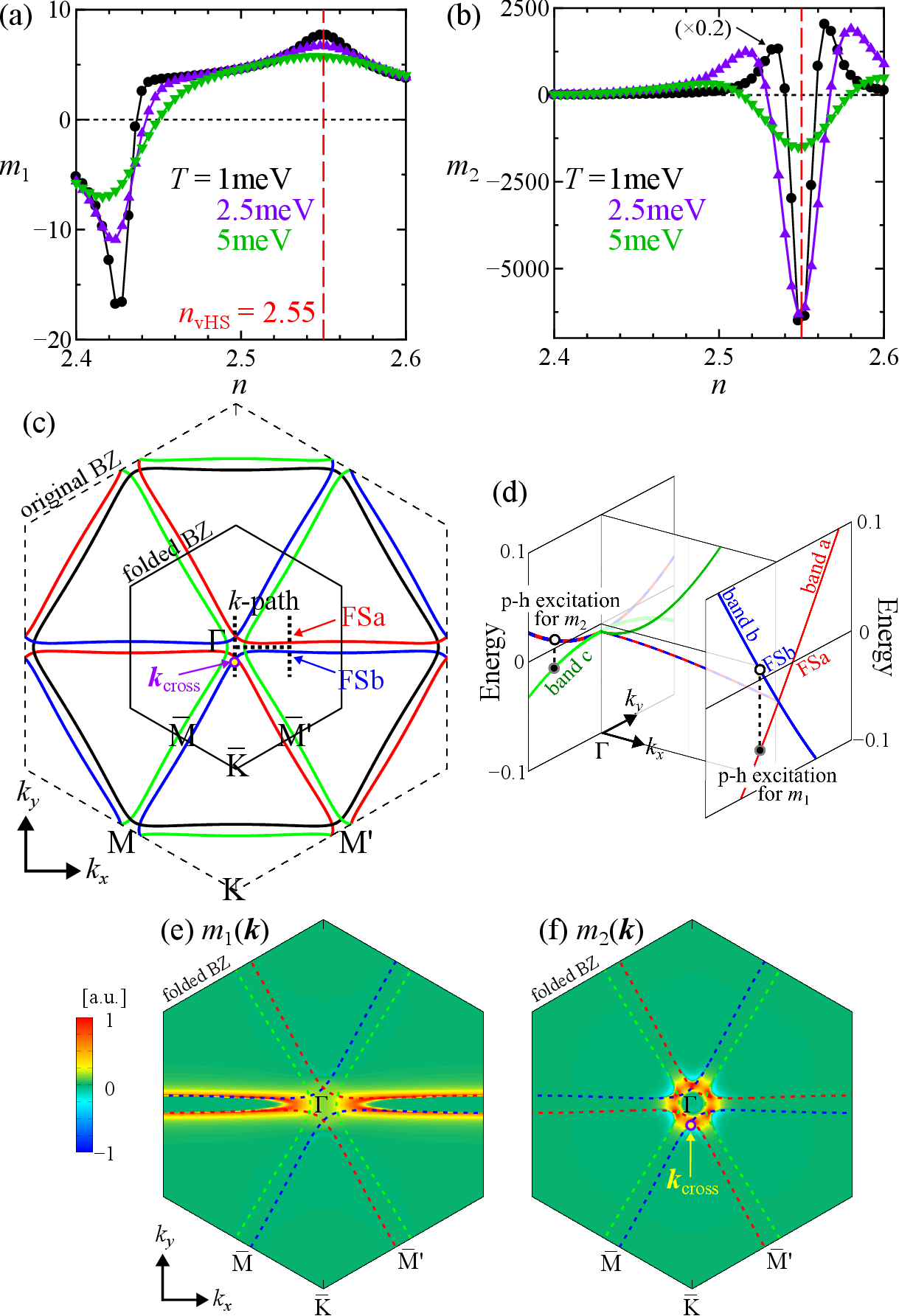}
\caption{
Obtained coefficients of ${\bar M}_{\rm orb}$,
(a) $m_1$ and (b) $m_2\ (\approx-m_3)$,
per $V$ atom as a function of $n$.
Both $|m_1|$ and $|m_2|$ are large for $n\sim n_{\rm vHS}$.
(c) Folded FS and (d) folded bandstructure in the folded BZ 
when ${\bm\eta}={\bm\phi}\rightarrow{\bm0}$.
The $\k$-path is shown in the (c).
We set the origin of the energy at the vHS energy.
(e) $m_1(\k)\equiv \d^2 m(\k)/\d\eta_1\d\phi_1|_{{\bm\eta}={\bm\phi}=0}$ and 
(f) $m_2(\k)\equiv \d^3 m(\k)/\d\eta_1\d\eta_2\d\eta_3|_{{\bm\eta}={\bm\phi}=0}$
for $n=2.58$ ($n_{\rm vHS}=2.55$).
Note that $m_n\propto\sum_\k m_n(\k)$.
}
\label{fig:fig3}
\end{figure}

Next stage, we calculate $M_{\rm orb}$
under the coexistence of the $3Q$ current order
${\bm \eta}=(\eta,\eta,\eta)/\sqrt{3}$
and the $3Q$ bond order
${\bm \phi}=(\phi,\phi,\phi)/\sqrt{3}$.
Figure \ref{fig:fig2} (d) represents $M_{\rm orb}$ 
at $\phi=0,\pm0.01$ as a function of $\eta$, for $n=2.47$ and $T=1$ meV.
Interesting relation $M_{\rm orb}\propto \eta$ is obtained when $\phi\ne0$.
Then, the field-induced $\Delta F$ contains a non-analytic $\eta$-linear term 
that always produces $\langle \eta \rangle\ne0$.
This fact causes significant field-induced change in the phase diagram,
as we will explain in the SI C \cite{SM}.

To understand the $\eta$-linear term in Fig. \ref{fig:fig2} (d),
we expand $M_{\rm orb}$ with respect to the current order
$\eta_m {\hat f}^{(m)}$ and bond order $\phi_m {\hat g}^{(m)}$.
Its Taylor expansion is
$\sum_{pqrp'q'r'} b_{pqr}^{p'q'r'}(\eta_1)^p (\eta_2)^q (\eta_3)^r (\phi_1)^{p'} (\phi_2)^{q'} (\phi_3)^{r'}$
with $p+q+r={\rm odd}$ and
$(p+p')\q_1+ (q+q')\q_2+(r+r')\q_3={\bm 0}$ (modulo original reciprocal vectors).
The $\eta$-linear term in
Fig. \ref{fig:fig2} (c) mainly originates from the second-order term
$b_{100}^{100}$ in addition to the third-order term $b_{100}^{011}$.
In fact, the $1Q$ current + $1Q$ BO state shown in Fig. \ref{fig:fig2} (e) 
violates the TRS in the bulk state.

\section*{Field-induced GL free energy expression}

From the above discussions,
we obtain the following convenient expression
of the orbital magnetization and the 
``field-induced GL free energy formula'' up to the third-order:
\begin{eqnarray}
{\bar M}_{\rm orb}&=& m_1 {\bm \phi}\cdot{\bm \eta} + m_2 \eta_1\eta_2\eta_3
\nonumber \\
& &+m_3(\eta_1\phi_2\phi_3 +\phi_1\eta_2\phi_3 +\phi_1\phi_2\eta_3)
\label{eqn:M-exp}, \\
\Delta {\bar F}&=&- 3h_z {\bar M}_{\rm orb} ,
\label{eqn:DeltaF2}
\end{eqnarray}
which enables us to elucidate the field-induced bond+current
phase transitions in kagome metals.
Figures \ref{fig:fig3} (a) and (b) show the coefficients $m_1$ and $m_2$ 
derived from Eq. [\ref{eqn:Mch}] numerically, respectively.
Interestingly, $|m_1|$ and $|m_2|$ become large for $n\sim n_{\rm vHS}$.

In the SI D \cite{SM}, 
we show that ${\bar M}_{\rm orb}$ in Eq. [\ref{eqn:M-exp}]
well reproduces the original $M_{\rm orb}$ 
when $|{\bm \eta}|,|{\bm \phi}|\lesssim 0.02$.
The expression of ${\bar M}_{\rm orb}$ 
is also justified based on the first principles model 
for $|{\bm \eta}|,|{\bm \phi}|\lesssim 0.02$ in the SI D \cite{SM}.

\textcolor{black}{
To understand the $n$-dependences of $m_1$ and $m_2$,
we discuss the folded FS and bandstructure 
given in Figs. \ref{fig:fig3} (c) and (d), respectively, 
for ${\bm\eta}={\bm\phi}\rightarrow{\bm0}$.
Here, band $a$ [$b$, and $c$] 
originates from the bandstructure around vHS-A [B,C]
in Fig. \ref{fig:fig1} (b).
As shown in Fig. \ref{fig:fig3} (d),
band $a$ and band $b$ degenerate along $k_x$-axis and $k_y$-axis.
Hereafter, we set $\mu=0$.
}

\textcolor{black}{
First, we consider the origin of $m_1$,
which is caused by the band-folding induced by 
$\eta_1{\hat f}^{(1)}$ and $\phi_1{\hat g}^{(1)}$,
both of which convey the wavevector $\q=\q_1$.
Therefore, $m_1$ is induced by the ``vertical p-h excitation 
between band $a$ and band $b$'' in Fig. \ref{fig:fig3} (d)
in the folded BZ.
(Band $c$ gives no contribution for $m_1$.)
To verify this consideration,
we examine the function $m(\k)$ in Eq. [\ref{eqn:Mch-integ}].
Figure \ref{fig:fig1} (e) shows the obtained 
$m_1(\k)\equiv \d^2 m(\k)/\d\eta_1\d\phi_1|_{{\bm\eta}={\bm\phi}=0}$.
(Note that $m_1\propto \sum_{\k}m_1(\k)$.)
It is verified that large $m_1$ originates from the 
FS $a$ and FS $b$ near the $\Gamma$-M line in Fig. \ref{fig:fig3} (c).
This mechanism occurs for both $n>n_{\rm vHS}$ and $n<n_{\rm vHS}$,
so $|m_1|$ takes large value for a wide range of $n$.
Note that $m_1$ in Fig. \ref{fig:fig3} (a) changes its sign at $n=2.44$,
when $\k_{\rm AB}$ and $\k'_{\rm AB}$ in Fig.\ref{fig:fig1} (b) 
touch in the folded BZ.
}

\textcolor{black}{
Next, we consider the origin of $m_2$,
which is caused by the band-foldings caused by 
${\hat f}^{(1)}$ (at $\q=\q_1$),
${\hat f}^{(2)}$ (at $\q=\q_2$) and ${\hat f}^{(3)}$ (at $\q=\q_3$).
This situation allows the ``vertical p-h excitation between band 
$a+b$ and band $c$'' in Fig. \ref{fig:fig3} (d),
which is significant because the band splitting 
between $\e_{c\k}\ (>0)$ and $\e_{a,b\k}\ (<0)$ is very small.
This process gives huge $|m_2|$ at $n= n_{\rm vHS}$
obtained in Fig. \ref{fig:fig3} (b).
Figure \ref{fig:fig3} (f) shows the obtained 
$m_2(\k)\equiv \d^3 m(\k)/\d\eta_1\d\eta_2\d\eta_3|_{{\bm\eta}={\bm\phi}=0}$.
(Note that $m_2\propto \sum_\k m_2(\k)$.)
The large $m_2$ originates from 
the six band-crossing points ($\k_{\rm cross}$) near the $\Gamma$ point,
due to the vertical p-h excitation among three bands $a$, $b$, $c$.
}

To summarize, large $|m_1|$ and $|m_2|$ are caused by the 
FS crossing near the vHS points and the M-M' line
for $n\sim n_{\rm vHS}$.
Therefore, the field-induced free energy $\Delta {\bar F}$ 
should be significant in real kagome metals.
In the SI B \cite{SM}, we verify that
the relation $m_3\approx-m_2$ holds very well.
This relation originates from the relation
$f_{ij}^{(m)}=ig_{ij}^{(m)}$ or $-ig_{ij}^{(m)}$
when both order parameters are composed of only the nearest bonds.

\textcolor{black}{
We verified in the SI E \cite{SM}
that the magnitudes of $m_1$ and $m_2$ 
at $t'=-0.08$ is comparable to those in Fig. \ref{fig:fig3}.
When $t'=-0.02$ in Fig. \ref{fig:fig3} (b),
the obtained $m_2(n_{\rm vHS})$ takes a large value 
because the p-h asymmetry around $E=E_{\rm vHS}$ is significant. 
When $t'=-0.08$ in Fig. S6 (e), 
in contrast, $m_2(n_{\rm vHS})$ is small and $m_2(n)$ 
tends to be an odd function of $n-n_{\rm vHS}$ 
because of the approximate p-h symmetry 
and the factor $(\e_{\a\k}+\e_{\b\k}-2\mu)$ in Eq. [\ref{eqn:MchT0}].
(The numerical results of $m_2(n)$ in the 30-orbital model 
in Figs. \ref{fig:fig6} (g)-(h) are closer to the results at $t'=-0.02$
near the vHS filling.)
}



Below, we derive the order parameters under $h_z$
by minimizing the GL free energy $F=F^0+\Delta {\bar F}$, where
$F^0=F^0_\eta+F^0_\phi+F^0_{\eta,\phi}$ is the free energy at $h_z=0$
\cite{Balents2021,Tazai-kagome2}:
%
\begin{eqnarray}
&&\!\!\!\!\!\!\!\!\!
F^0_\phi= a_{\rm b} |{\bm \phi}|^2+ b_1\phi_1\phi_2\phi_3 
\nonumber\\
& &+d_1 (\phi_1^4+\phi_2^4+\phi_3^4)+ d_2 (\phi_1^2\phi_2^2+({\rm cycl.})) ,
\label{eqn:Fphi}
\\
&&\!\!\!\!\!\!\!\!\!
F^0_\eta= a_{\rm c} |{\bm \eta}|^2+ d_3 (\eta_1^4+\eta_2^4+\eta_3^4)+ d_4 (\eta_1^2\eta_2^2+({\rm cycl.})),
\label{eqn:Feta}
\end{eqnarray}
and $F^0_{\eta,\phi}$ contains the current-bond cross terms
proportional to $\eta^2\phi^1$ and $\eta^2\phi^2$:
\begin{eqnarray}
&&\!\!\!\!\!\!\!\!\!
F^0_{\eta,\phi}= b_2(\phi_1\eta_2\eta_3 + ({\rm cycl.}))
\nonumber\\
& &+2d_5(\phi_1^2\eta_1^2+\phi_2^2\eta_2^2+\phi_3^2\eta_3^2)
+ d_6(\phi_1^2\eta_2^2+({\rm cycl.})).
\label{eqn:Fphi-eta}
\end{eqnarray}
Here, $a_{\rm c(b)}=r_{\rm c(b)}(T-T_{\rm c(b)}^0)$,
where $T_{\rm c(b)}^0$ is the current-order (BO) 
transition temperature without other orders.
Theoretically,
$a_{\rm c(b)}\sim N(0)(-1+\lambda_{\eta(\phi)}^{-1})$,
where $N(0)$ is the density-of-states ($\sim1 \ {\rm eV^{-1}}$)
and $\lambda_{\eta(\phi)}$ is the DW equation eigenvalue 
of the current order (bond order) \cite{Tazai-LW}.
$\lambda_{\eta(\phi)}=1$ at $T=T_{\rm c(b)}^0$,
while $\lambda_{\eta(\phi)}=0$ ({\it i.e.}, $a_{\rm c(b)}=\infty$)
in the absence of interaction.
According to Ref. \cite{Tazai-LW},
$r_{\rm b}T_{\rm b}^0\sim 0.1N(0)$ for the nematic BO in FeSe,
while the corresponding value for BCS superconductors
is $\sim N(0)$.

To discuss the phase diagram qualitatively,
we set $m_1=5$, $m_2=-m_3=-1000$,
both of which are moderate compared with the values
in Fig. \ref{fig:fig3}.
We also put $d_i=150$ $(i=1-4)$
by seeing the numerical results in the SI F \cite{SM}
and set $r_{\rm c}=r_{\rm b}=30$.
Then, both current and bond orders become $3Q$ states
because of the relations $2d_1/d_2>1$ and $2d_3/d_4>1$.
A more detailed explanation for the GL parameters
is presented in the SI F \cite{SM}.


\section*{$h_z$-effect on $3Q$ current order state}

\begin{figure}[htb]
\includegraphics[width=.99\linewidth]{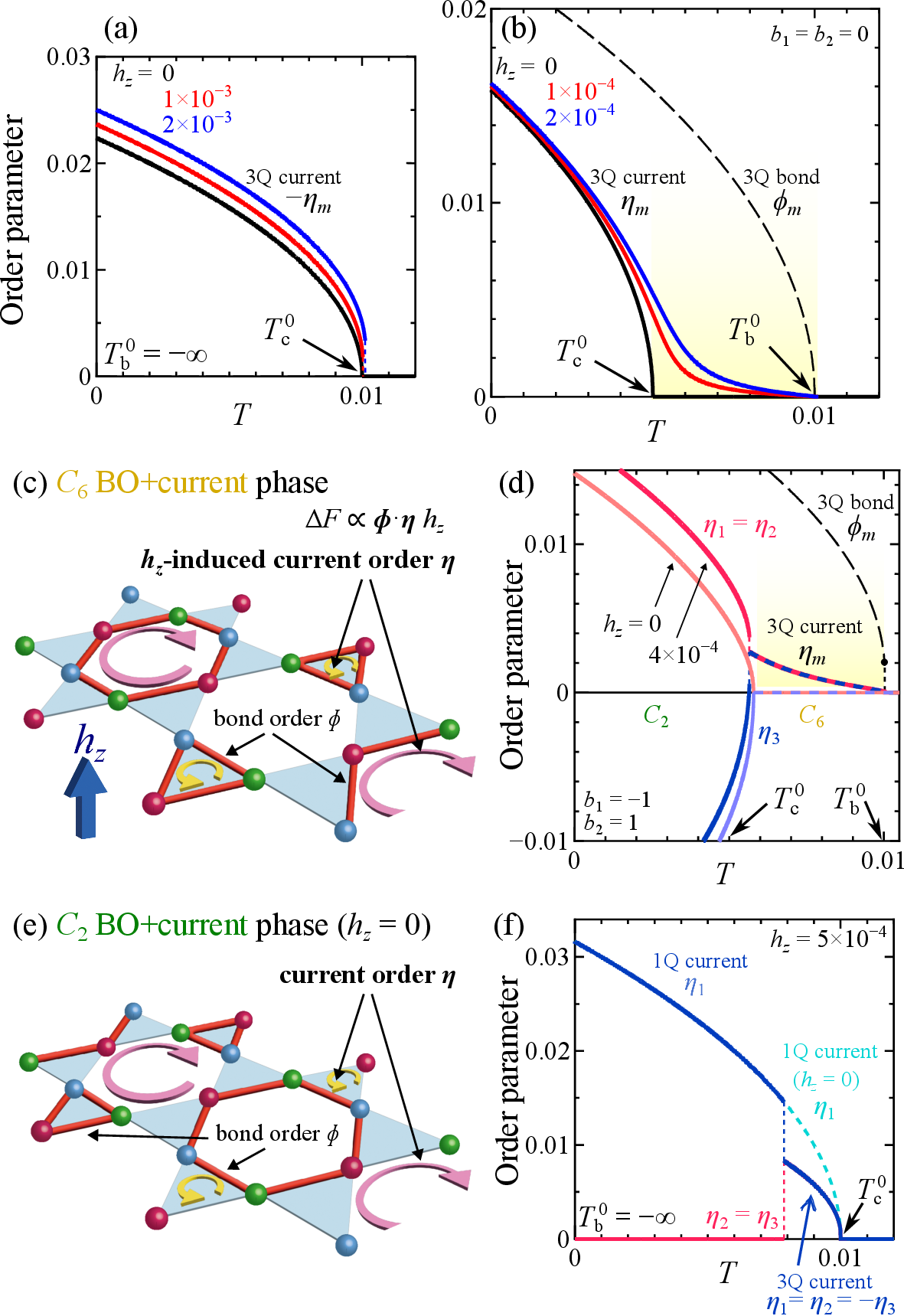}
\caption{
(a)(b) Obtained $3Q$ current order 
${\bm\eta}=(\eta_1,\eta_2,\eta_3)$ with $\eta_1=\eta_2=\eta_3$ 
for $2d_1/d_2=2$, $2d_3/d_4=2$ and $b_1=b_2=0$ under $h_z$.
$h_z=10^{-4}$ corresponds to 1 Tesla.
(a) $3Q$ current order for $T_{\rm c}^0=0.01$ and $T_{\rm b}^0=-\infty$.
The first-order transition is induced by $h_z$ due to $m_2$-term.
(b) $3Q$ current order for $T_{\rm c}^0=0.005$
under the $3Q$ BO ${\bm\phi}\propto(1,1,1)$ with $T_{\rm b}^0=0.01$.
For $h_z\ne0$,
$\eta$ starts to increase below $T_{\rm b}$ thanks to $m_1$-term,
{\it i.e.}, $T_{\rm c}=T_{\rm b}^0$.
(c) $h_z$-induced $C_6$-symmetry bond+current order 
above $T_{\rm c}^0$ the shaded area in (b).
(d) $3Q$ current order for $-b_1=b_2=1.0$.
Other model parameters are equal to (b). 
The symmetry of the bond+current coexisting state is 
$C_6$ ($\eta_1=\eta_2=\eta_3$) for $T>T_{\rm c}^0$, while 
$C_c$ ($\eta_1=\eta_2\approx-\eta_3$) for $<T_{\rm c}^0$.
(e) Bond+current coexisting state below $T_{\rm b}^0$.
It has $C_2$ symmetry to gain the 3rd order GL terms.
(f) Obtained current order $\eta_m$ for $2d_3/d_4=1/1.2$
at $T_{\rm c}^0=0.01$ and $h_z=5\times10^{-4}$, for $T_{\rm b}^0=-\infty$.
The $3Q$ current order appears due to the $m_2$-term,
while it changes to $1Q$ order at $T\approx0.008$.
}
\label{fig:fig4}
\end{figure}

In this section, we demonstrate that the $3Q$ current order state
is drastically modified by $h_z$.
Figure \ref{fig:fig4} (a) shows the 
obtained $3Q$ current order ${\bm \eta}=(\eta_1,\eta_2,\eta_3)$ 
with $\eta_1=\eta_2=\eta_3$, when $T_{\rm c}^0=0.01$
in the absence of BO ($T_{\rm b}^0=-\infty$).
Here, we set $h_z=0\sim2\times10^{-3}$.
Because of the $\eta^3$-term by $m_2=-1000$,
$3Q$ current order ${\bm \eta}=(\eta,\eta,\eta)/\sqrt{3}$
with negative $\eta$ is stabilized in the presence of $h_z>0$.
In addition, the field-induced first-order transition
occurs at $T=T_{\rm c}^0+\Delta T_{\rm c}$,
where $\Delta T_{\rm c}=(h_zm_2)^2/4r_{\rm c}(d_3+d_4)$ is quite small.

\textcolor{black}{
Next, we demonstrate the $h_z$-induced $3Q$ current order state 
above $T_{\rm c}^0$ inside the BO state.
Under the $3Q$ BO phase ${\bm\phi}=(\phi,\phi,\phi)/\sqrt{3}$,
the 2nd order GL coefficient $a_{\rm c}$ in Eq. [\ref{eqn:Feta}] is renormalized 
as ${\bar a}_{\rm c}=a_{\rm c}+(d_5+d_6)(2\phi^2/3)$
due to the $d_5$, $d_6$ terms.
It is given as ${\bar a}_{\rm c}={\bar r}_{\rm c}(T-{\bar T}_{\rm c}^0)$,
where ${\bar T}_{\rm c}^0<T_{\rm c}^0$ and ${\bar r}_{\rm c}<r_{\rm c}$ 
as we explain in the SI F \cite{SM}.
The smallness of ${\bar r}_{\rm c}$ is favorable for the 
$h_z$-induced current order.
In this subsection, we simply denote ${\bar T}_{\rm c}^0$ and ${\bar r}_{\rm c}$ 
as $T_{\rm c}^0$ and $r_{\rm c}$, respectively.
}

\textcolor{black}{
First, we drop the 3rd order GL terms $b_i$ ($i=1,2$)
to concentrate on the field-induced novel phenomena.
Figure \ref{fig:fig4} (b) exhibits the 
$3Q$ current order with $T_{\rm c}^0=0.005$,
in the presence of the BO phase below $T_{\rm b}^0=0.01$.
For $h_z=1\times 10^{-4}$ and $2\times 10^{-4}$,
$\eta$ starts to emerge just below $T_{\rm b}^0$
thanks to the $\eta$-linear term 
$\Delta {\bar F} = -3h_z(m_1\phi+m_3\phi^2)\eta$,
and therefore $T_{\rm c}=T_{\rm b}^0$.
When $h_z\ne0$, $T_{\rm c}^0$ is just a crossover temperature.
To understand the role of the non-analytic term qualitatively,
we analyze a simple GL free energy with a $\eta$-linear term
in the SI C \cite{SM}:
It is found that the field-induced current order $|{\bm\eta}_{{\rm ind}}|$
under the BO ${\bm\phi}$ is approximately given as
\begin{eqnarray}
|{\bm\eta}_{{\rm ind}}|\approx 9|m_1 h_z {\bm\phi}|/2a_{\rm c}.
\label{eqn:hind}
\end{eqnarray}
Thus, the field-induced $|{\bm\eta}_{{\rm ind}}|$ is prominent
when the system at $h_z=0$ is close to the current order state
({\it i.e.}, $a_{\rm c}\gtrsim0$).
The expression of $|{\bm\eta}_{{\rm ind}}|$ in Eq. [\ref{eqn:hind}]
is proportional to ${\bm\phi}$
in contrast to Eq. (31) of Ref. \cite{Balents2021}.
}

\textcolor{black}{
A schematic picture for the field-induced current order
is shown in Fig. \ref{fig:fig4} (c).
In the $3Q$ BO phase ${\bm \phi}\propto(1,1,1)$,
the field-induced ${\bm\eta}$ above $T_{\rm c}^0$ is proportional to 
${\bm\phi}$ to maximize ${\bar M}_{\rm orb}\propto{\bm\phi}\cdot{\bm\eta}$.
This coexisting state (${\bm\eta}\propto{\bm\phi}$) 
has $C_6$ symmetry; see Ref. \cite{Tazai-kagome2}.
}

\textcolor{black}{
Next, we consider the effect of 3rd order GL terms in Fig. \ref{fig:fig4} (d),
by setting $-b_1=b_2=1.0$.
(The relation $b_1b_2<0$ is general \cite{Tazai-kagome2}.)
Due to the energy gain from the $b_1$-term,
3Q BO ${\bm \phi}\propto(1,1,1)$
appears at $T\approx T_{\rm b}^0=0.01$ as the first order transition
\cite{Balents2021}.
At $h_z=0$, the 3Q current order ${\bm \eta}=(\eta,\eta,-\eta')$ 
appears below $T_{\rm c}^0=0.005$
to maximize the energy gain from the $b_2$ term ($b_1b_2<0$),
as explained in Ref. \cite{Tazai-kagome2}.
Just below $T_{\rm c}^0$ ($|{\bm \phi}|\gg|{\bm \eta}|$),
$\eta'=2\eta$ is satisfied.
(More generally, ${\bm \eta}\propto(\eta_1,\eta_2,-\eta_1-\eta_2)$.)
Figure \ref{fig:fig4} (e) depicts the
``nematic'' bond+current coexisting state below $T_{\rm c}^0$
with ${\bm\eta}\sim(1,1,-1)$ and ${\bm\phi}\sim(1,1,1)$;
see Ref. \cite{Tazai-kagome2} for detail.
}

\textcolor{black}{
For $h_z=4\times 10^{-4}$ in Fig. \ref{fig:fig4} (d),
the field-induced $3Q$ current order 
${\bm \eta}\propto(1,1,1)$ start to emerge just below $T_{\rm b}^0$,
similarly to Fig. \ref{fig:fig4} (b).
The realized $C_6$ symmetry coexisting state above $T_{\rm c}^0$ is 
shown in Fig. \ref{fig:fig4} (c).
Below $T_{\rm c}^0$, the coexisting state changes to the 
nematic ($C_2$) bond+current state shown in Figure \ref{fig:fig4} (e).
We stress that ${\rm sgn}(\eta_1\eta_2\eta_3)$ changes at $T\approx T_{\rm c}^0$. 
To summarize, the $h_z$-induced coexisting state changes its symmetry 
from $C_6$ ($T>T_{\rm c}^0$) to $C_2$ ($T<T_{\rm c}^0$) with decreasing $T$.
The field-induced first-order phase transition occurs at $T\approx T_{\rm c}^0)$.
}



\section*{$h_z$-effect on $1Q$ current order state}
In this scetion, we explain that the $1Q$ current order state
is also drastically modified by $h_z$.
Here, we increase $d_4=150$ to $360$
({\it i.e.}, $2d_3/d_4=1/1.2$) in the previous GL parameters.
A clear evidence of the $1Q$ current order 
has been reported recently in Ref. \cite{Asaba}.
We set $T_{\rm c}^0=0.01$.
Figure \ref{fig:fig4} (f) shows the obtained order parameters
without BO ($T_{\rm b}^0=-\infty$) at $h_z=5\times10^{-4}$.
We obtain the field-induced $3Q$-current order due to the $m_2$-term,
while it changes to $1Q$ order at $T\approx0.008$.
The field-induced first-order transitions 
reported in Ref. \cite{Asaba}
would originate from the $1Q$ current order at $T^{**} \ (>T_{\rm b}^0)$
together with $\Delta {\bar F}$.
In Fig. \ref{fig:fig4} (f), we set $2d_3/d_4=1.2$ and $m_2=1000$.
The obtained $h_z$-induced $3Q$ current order is realized
even when $m_2\approx100$ when $2d_3/d_4\gtrsim1$.

\textcolor{black}{
In Fig. \ref{fig:fig4} (f), we set $T_{\rm b}^0=-\infty$ for simplicity.
However, $T_{\rm b}^0\approx0.01 \ (\approx100\ {\rm K})$ experimentally.
In this case, as revealed in Ref. \cite{Balents2021},
the $3Q$ current order induces the 
finite $3Q$ BO even above $T_{\rm b}^0$ via the $b_2$ term.
This fact means that the $3Q$ current order is energetically favorable 
when $b_2\ne0$.
Therefore, $h_z$-induced transition from $1Q$ to $3Q$ current order 
shown in Fig. \ref{fig:fig4} (f) can emerge even if $2d_3/d_4\lesssim1$
for $b_2\ne0$.
}

For reference,
we studied the case of the intra-original-unit-cell current order
($\q={\bm 0}$) in kagome lattice in the SI G \cite{SM}.
In this case, $M_{\rm orb}$ is $\eta$-linear even if the BO is absent.

%

\section*{Derivation of GL coefficients based on the first-priciples kagome metal model}

\textcolor{black}{
In the next stage, we calculate the GL coefficients 
based on the first-principles tight-binding model for kagome metals.
We reveal that $|m_1|$ and $|m_2|$ becomes large
due to the ``inter-orbital ($d_{XZ}+d_{YZ}$) mixture''
even if the current order emerges only in the $d_{XZ}$-orbital.
}

\begin{figure}[htb]
\includegraphics[width=.99\linewidth]{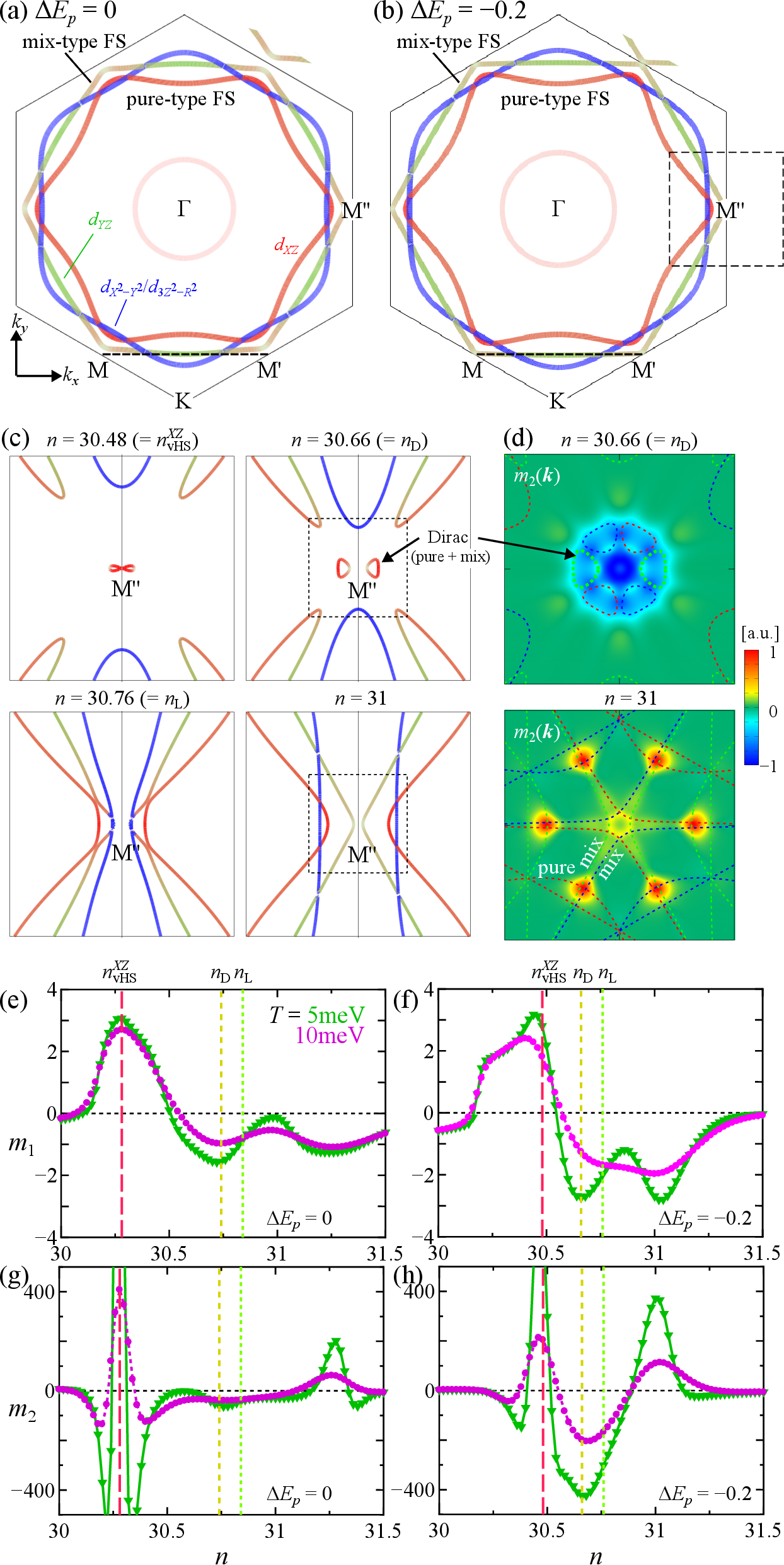}
\caption{
(a) FS of CsV$_3$Sb$_5$ model in the $k_z=0$ plane.
The $d_{XZ}$-orbital band gives the pure-type FS,
and the $d_{YZ}$-orbital band gives the mix-type FS.
The $d_{X^2-Y^2}$+$d_{3Z^2-R^2}$-orbital band gives another pure-type FS.
(b) FS with the $p$-orbital shift $\Delta E_{p}=-0.2$ eV.
Here, the mix-type FS is closer to the M points.
(c) Change in the FS topology with $n$ in $E_p=-0.2$ model
inside the dashed line square in (b)
(d) $m_2(\k)$ inside the dotted line square in (c).
with the folded FSs at $n=n_{\rm D}$ and $n=31$.
(e-h) Obtained $m_n$ at $T=0.01$ and $0.005$:
(e) $m_1$ in $\Delta E_{p}=0$ model, where $m_1\sim -0.5$ at $n=31$.
(f) $m_1$ in $\Delta E_{p}=-0.2$ model, where $m_1\sim -2$ at $n=31$.
(g) $m_2$ in $\Delta E_{p}=0$ model, where $m_2\sim-50$ at $n=31$.
(h) $m_2$ in $\Delta E_{p}=-0.2$ model, where $m_2\sim300$ at $n=31$.
}
\label{fig:fig6}
\end{figure}

\textcolor{black}{
First, we derive 30-orbital (15 $d$-orbital and 15 $p$-orbitals)
tight-binding model for CsV$_3$Sb$_5$ by using 
WIEN2k and Wannier90 softwares.
Figure \ref{fig:fig6} (a) shows the FS of the obtained model
in the $k_z=0$ plane.
(Its bandstructure is shown in Fig. S9 (a).)
The $d_{XZ}$-orbital ``pure-type'' FS
corresponds to the FS of the present three-orbital model.
Its vHS energy is located at $E_{\rm vHS}^{XZ}\approx-0.1$.
Also, the $d_{YZ}$-orbital forms the ``mix-type'' FS
whose vHS energy is $E_{\rm vHS}^{YZ}\approx+0.1$.
In addition, both $d_{X^2-Y^2}$ and $d_{3Z^2-R^2}$ orbitals form one 
pure-type band with the vHS energy $E_{\rm vHS}^{X^2-Y^2}\approx0$.
}

\textcolor{black}{
Figure \ref{fig:fig6} (b) shows the FS 
with introducing the $p$-orbital shift $\Delta E_{p}=-0.2$ eV.
Its bandstructure is shown in Fig. S9 (b).
Here, the $d_{YZ}$-orbital (mix-type) FS approaches the M point, 
and its vHS point shifts to the Fermi level,
consistently with the ARPES measurement in Ref.
\cite{ARPES-VHS}.
Figure \ref{fig:fig6} (c) shows the change in the FS topology 
due to the pure-mix hybridization in the $E_p=-0.2$ model for 
$n=30.48\sim 31.00$.
When $n=n_{\rm vHS}^{XZ}\ (=30.48)$, 
we obtain two hole-pockets that attach the M point.
With increasing $n$,
it changes to the electron-like Dirac pockets at $n\approx n_{\rm D} \ (=30.66)$,
composed of both $d_{XZ}$ and $d_{YZ}$ orbitals.
At $n=n_{\rm L}\ (=30.76)$, 
large hole-like pure-type FS and large electron-like mix-type FS are formed.
At $n=31$, the mix-type FS lines on the M-M' line.
We will see that large $|m_1|$ and $|m_2|$ appear
at $n\approx n_{\rm vHS}$, $n_{\rm D}$ and $31$.
}

\textcolor{black}{
Figure \ref{fig:fig6} (d) shows the obtained $m_2(\k)$
with the folded FSs.
For both $n=n_{\rm D}$ and $31$, the pure-mix hybridization
contributes to the large GL coefficient in the 30-orbital model.
This mechanism is absent in the simple single orbital model.
}

\textcolor{black}{
We note that $E_{\rm vHS}^{XZ}\approx0$ is reported in 
an ARPES study for pristine CsV$_3$Sb$_5$ in Ref. \cite{Sato-ARPES}.
In this case, both $m_1$ and $m_2$ become very large theoretically.
}

\textcolor{black}{
Now, we calculate $m_1$ and $m_2$ in ${\bar M}_{\rm orb}$
by introducing the current and bond orders on the $d_{XZ}$ orbitals
in the 30-orbital kagome lattice model 
given in Figs. \ref{fig:fig6} (a) ($\Delta E_{p}=0$)
and (b) ($\Delta E_{p}=-0.2$).
We derive the coefficients $m_1$ and $m_2$ defined as
${\bar M}_{\rm orb}= m_1 {\tilde{\bm \phi}}\cdot{\tilde{\bm \eta}} + m_2 {\tilde{\eta_1}}{\tilde{\eta_2}}{\tilde{\eta_3}}$,
where $\tilde{\eta_i}$ and $\tilde{\phi_i}$ are the 
$d_{XZ}$-orbital order parameters projected on the pure-type band. 
(We set $\tilde{\eta_i}=\eta_i W_{XZ}$ and $\tilde{\phi_i}=\phi_i W_{XZ}$
with $W_{XZ}=0.7$, which is the $d_{XZ}$-weight on the pure-type band.)
}

\textcolor{black}{
The obtained $m_1$ is shown in Figs. \ref{fig:fig6} (e) ($\Delta E_{p}=0$) 
and (f) ($\Delta E_{p}=-0.2$),
as function of the electron filling $n$.
Here, $m_{1(2)}$ is derived from the $k_z=0$ plane electronic structure.
First, we discuss the case $n=31$ that corresponds to undoped CsV$_3$Sb$_5$.
In both Figs. \ref{fig:fig6} (e) and (f),
the obtained $m_1$ is very larger for $n\approx n_{\rm vHS}^{XZ}\ (<31)$.
At $n=31$, $m_1$ in Fig. \ref{fig:fig6} (e) is relatively small 
($\sim-0.5$) accidentally, while its magnitude becomes large ($m_1\sim-2$)
in Fig. \ref{fig:fig6} (f), where the mix-type FS is closer to the M-M' line.
In fact, the mix-type FS contains finite $d_{XZ}$-weight ($\sim10$\%)
owing to the inter-orbital mixture in kagome metal bandstructure.
For this reason, the mix-type band can cause large $M_{\rm orb}$
even if the current order parameter occurs only in $d_{XZ}$-orbital.
Thus, the current-bond-field trillinear coupling 
due to $m_1$ term will cause the drastic field-induced chiral current order 
shown in Fig. \ref{fig:fig4} (b).
This is the main result of this study.
}

\textcolor{black}{
The obtained $m_2$ is shown in Figs. \ref{fig:fig6} (g) ($\Delta E_{p}=0$) 
and (h) ($\Delta E_{p}=-0.2$), at $T=0.01$ and $0.005$.
The obtained $m_2$ is extremely larger for $n\approx n_{\rm vHS}^{XZ}\ (<31)$.
At $n=31$, 
$m_2$ becomes relatively small ($\sim-50$) in Fig. \ref{fig:fig6} (g).
However, $m_2$ increases to $\sim400$ in Fig. \ref{fig:fig6} (h)
because the mix-type FS is closer to the M-M' line.
Thus, large $|m_2|$ can be realized by the mix-type band 
with finite $d_{XZ}$-orbital weight
even when $|E_{\rm vHS}^{XZ}-\mu|\sim0.1$.
We stress that the relation $m_2\approx -m_3$ 
is well satisfied; see the SI B \cite{SM}.
}

\clearpage

\textcolor{black}{
Next, we discuss the $n$-dependences of $m_1$ and $m_2$ in more detail.
In Figs. \ref{fig:fig6} (e)-(h), 
both coefficients take very large values at 
$n\approx n_{\rm vHS}^{XZ}$ due to the 
hybridization between $d_{XZ}$-orbital bands,
as we explained in Fig. \ref{fig:fig3}.
With increasing $n$,
these coefficients exhibit drastic $n$-dependences
when the FS changes its topology due to the pure-mix hybridization.
When $n\approx n_{\rm D}$, 
the electron-like Dirac pockets made of $d_{XZ}+d_{YZ}$ orbitals 
cause large $|m_1|$ and $|m_2|$.
$|m_1|$ and $|m_2|$ are also enlarged when
the mix-band (with finite $d_{XZ}$-weight) 
is close to the M-M' line
at $n\approx31$ in $\Delta E_{p}=-0.2$ model
[at $n\approx31.2$ in $\Delta E_{p}=0$ model].
}

\textcolor{black}{
To summarize,
both $m_1$ and $m_2$ exhibit interesting sensitivity to the 
multiorbital character and the nesting condition 
of the bandstructure around M points.
Unexpectedly, both $d_{XZ}$-orbital (pure-type) band and 
$d_{YZ}$-orbital (mix-type) band play important role for $m_1$ and $m_2$ 
even when the current order emerges only on the $d_{XZ}$-orbital.
When $\Delta E_{p}=-0.2$, the mix-type FS approaches the M point,
consistently with the ARPES bandstructure \cite{ARPES-VHS}.
Then, we obtain large $|m_1|(\sim2)$ and $|m_2|(\lesssim400)$ 
at $n=31$ as shown in Figs. \ref{fig:fig6} (f) and (h).
Thus, the $h_z$-induced change in Fig. \ref{fig:fig4}
can be obtained by changing the parameters slightly.
(``Nearly commensurate'' current order may appear in $\Delta E_{p}=0$ model,
in which the nesting of the FS at $\q_m$ is not perfect \cite{Tazai-kagome}.
The calculated $m_n$ will be larger by using the correct 
incommensurate nesting vector of $\Delta E_{p}=0$ model.
)
In the SI H \cite{SM},
we calculate the GL coefficients 
when the current and bond orders emerge on other $d$-orbitals.
It is found that large $m_1$ and $m_2$ are obtained in various cases.
Thus, the present study is valid for various types of 
current order mechanisms, not restricted to 
Ref. \cite{Tazai-kagome2}.
}

\section*{Summary and Discussions}

\textcolor{black}{
In summary, the 
chiral current order in kagome metal exhibits weak ferromagnetism,
and its magnitude are enlarged in the bond order state.
Importantly, 
we derived the $h_z$-induced GL free energy expression
$\Delta {\bar F}$ in Eq. [\ref{eqn:DeltaF2}],
which provides an important basis for future research in kagome metals.
The emergent $\eta\cdot\phi\cdot h_z$ trilinear term in $\Delta {\bar F}$
explains prominent field-induced chiral symmetry breaking in kagome metals.
We revealed that 
$m_n$ becomes large when the vHS energy is close to the Fermi level because the prominent FS reconstruction occurs due to the current order parameters.
Furthermore, the multi-orbital mixing in the bandstructure of real kagome metals makes $m_n$ larger for a wide filling range.
The finding that $m_n$ sensitively depends on the multiorbital bandstructure in Fig. \ref{fig:fig6} provides a useful hint to control the charge current or to understand the difference between Cs-based and K-based compounds \cite{Moll-K}. 
Interestingly, we obtain large $m_n$ when the current order parameter emerges in not only $d_{XZ}$-orbital, but also other $d$-orbitals.
It is interesting to study $m_n$ for theoretically proposed
exotic TRS breaking states, such as the inter-orbital exciton order
\cite{Nat-g-ology}
and multipolar and toroidal magnetic orders
\cite{Fernandes-GL}.
}

Below, we discuss several important issues.

\textcolor{black}{
\subsection{Comparison with experiments: $h_z$--induced current order}\ 
In Ref. \cite{Moll-hz}, it was proposed that
CsV$_3$Sb$_5$ is located at the quantum critical point of 
the current order (${\bar T}_{\rm c}^0\approx0$) 
in the absence of the uniaxial strain.
(As we discussed above, $T_{\rm c}^0$ is renormalized to ${\bar T}_{\rm c}^0$ under the BO phase.)
The field-induced ($h_z\sim9$T) current order at $T\sim20$K
is naturally understood based on the GL free energy analysis
with the current-bond-$h_z$ trillinear term.
Figures \ref{fig:fig4} (b) and (d) corresponds to the experimental report in
Ref. \cite{Moll-hz} by considering that ${\bar T}_{\rm c}^0\approx0$.
In addition, the present GL theroy explains the field-induced 
enhancement of the local magnetic flux ($\propto \eta^1$) 
observed by $\mu$SR measurements in AV$_3$Sb$_5$
\cite{muSR4-Cs,muSR2-K,muSR5-Rb}
and field-tuned chiral transport study 
\cite{eMChA}.
Interestingly, the obtained $m_n$ sensitively 
depends on the multiorbital bandstructure and the filling
in Fig. \ref{fig:fig6} (and Fig. S9).
The present discovery provides a useful hint to control 
the charge current in kagome metals. 
The present finding will also be essential to understand 
the significant difference between Cs-based and K-based compounds 
reported in Ref. \cite{Moll-K}. 
In addition, the discovery of CDW state in double-layer kagome metal ScV$_6$Sn$_6$ \cite{166-CDW} has attracted increasing theoretical interest \cite{Thomale-GL,Bernevig-166-exp,Bernevig-166}.
The TRS breaking state and its increment under $h_z$ discovered in ScV$_6$Sn$_6$ \cite{166-mSR} may be understood by developing the present GL theory.
}

\textcolor{black}{
We note that the trilinear term in $\Delta {\bar F}$
is also caused by the spin magnetization in the presence of 
spin-orbit interaction
\cite{Fernandes-GL}.
Future quantitative calculations would be important.
}

\textcolor{black}{
\subsection{Comparison with experiments: Strain-induced current order}\ 
Reference \cite{Moll-hz} also reports interesting prominent
strain induced increment of ${\bar T}_{\rm c}^0$.
Under the uniaxial strain, the degeneracy of 
the current order transition temperature at $\q=\q_m$ ($m=1\sim3$),
${\bar T}_{\rm c}^{(m)}$, is lifted by the change in the 2nd order GL term
as discussed in Ref. \cite{Moll-hz}.
Then, ${\bar T}_{\rm c}= \max_{m} {\bar T}_{\rm c}^{(m)}$
will be larger than the original ${\bar T}_{\rm c}^0$.
In the SI I \cite{SM},
we find that additional significant contribution to the increment of 
${\bar T}_{\rm c}^0$
originates from the strain induced change in the 4th order GL terms ($d_i$).
Here, the ``BO-induced suppression of the current order by $d_5$ and 
$d_6$ GL terms is found to be drastically reduced by the strain.
In fact, we find that $d_i$ is sensitive to the strain
because the vHS energy is close to the Fermi level; 
see Fig. S10 in the SI I \cite{SM}.
}

\textcolor{black}{
Below, we explain the discussion in the SI I \cite{SM}.
Considering the $D_{6h}$ symmetry, 
we assume that the $E_{1g}$ symmetry strain $(\e,\e')$ 
induces the shift of the vHS energy levels as
$\Delta{\bm E}\equiv (\Delta E_A,\Delta E_B, \Delta E_C)= \a(\e,0,-\e)$.
($\Delta{\bm E}'$ by $\e'$ is presented in the SI I \cite{SM}.)
Hereafter, we set $\a=1$.
Under the $3Q$ BO phase ${\bm\phi}^0=(\phi^0,\phi^0,\phi^0)/\sqrt{3}$,
the 2nd order GL coefficient $a_{\rm c}$ in Eq. [\ref{eqn:Feta}]
in the absence of the strain ($\e=0$) is changed to 
\begin{eqnarray}
{\bar a}_{\rm c}(0)&=& a_{\rm c} + 2(d^5+d^6)((\phi^0)^2/3)
\nonumber \\
&\propto& T-{\bar T}_{\rm c}^0,
\label{eqn:ac0}
\end{eqnarray}
where $|\phi^0|\gtrsim T_{\rm b}$ when $T\ll T_{\rm b}$.
Here, we assume that $\phi^0$ is $T$-independent because we study the strain-induced current order for $T\ll T_{\rm b}^0$.
For finite $\e$, it is changed by $\Delta F(\e)$ as
\begin{eqnarray}
{\bar a}_{\rm c}(\e)&=& {\bar a}_{\rm c}^0-[(2g_5+g_6)+D(2g_1+g_2)]((\phi^0)^2/6) 
\nonumber \\
&\propto& T-{\bar T}_{\rm c}(\e),
\label{eqn:ac}
\end{eqnarray}
where $g_l$ are $\e$-linear.
Here, ${\bar T}_{\rm c}(0)={\bar T}_{\rm c}^0$.
$g_l$ and $d_l^{(m)}$ are related as
$\e\cdot ( \d_{\e} d_l^{(1)},\d_{\e} d_l^{(2)},\d_{\e} d_l^{(3)})=(-1)^l\cdot g_l(1,-1,0)$.
The the current order appears in the BO phase
when Eq. [\ref{eqn:ac}] becomes negative.
Therefore, ${\bar T}_{\rm c}(\e)$ will increase in proportion to $\e$ if $(2g_5+g_6)>0$.
Therefore, we obtain
${\bar T}_{\rm c}(\e)\sim (\e/\e_0) (T_{\rm c}^0-{\bar T}_{\rm c}^0)$
with $\e_0=0.025$ ($0.029$) for $n=2.56$ ($2.48$),
according to the numerical study in Fig. S10.
This increment of ${\bar T}_{\rm c}(\e)$ originates from the
reduction of the BO-induced suppression of the current order.
Therefore, 
considering that $T_{\rm c}^0\sim0.01$ and ${\bar T}_{\rm c}^0=0$ (= at current order QCP),
we obtain ${\bar T}_{\rm c}(\e)\sim0.002$ ($\approx20$ K) 
at $\e\sim0.005$ $(\approx 50\ {\rm K})$. 
Based on this extrinsic strain scenario,
one may understand the difference of $T_{\rm c}$ by experiments,
such as the absence of the TRS breaking at $h_z=0$ reported by Kerr rotation study \cite{Kapitulnik}.
}



\textcolor{black}{
\subsection{Nematic 3D stacking $3Q$ BO and current order}\ 
As revealed in Ref. \cite{Balents2021},
the nematic state can be realized by the 
$\pi$-shift 3D stacking of the $3Q$ BO thanks to the 3rd order GL term $b_1$.
This state is the combination of three BO states
at wavevectors $\q_l^{\rm 3D}=(\q_l^{\rm 2D},q_l^z)$ for $l=1,2,3$
(${\bm \q}_l^{\rm 2D}$ is shown in Fig. \ref{fig:fig1} (a))
when $\{q_1^z,q_2^z,q_3^z\}=\{\pi,\pi,0\}$, $\{\pi,0,\pi\}$ or $\{0,\pi,\pi\}$.
Because of the relation $\sum_l^{1,2,3}\q_l^{\rm 3D}={\bm0}$,
the $\pi$-shift stacking gains the 3rd order GL free energy
due to the $b_1$ term.
Its necessary condition is that the 2nd order GL term $a_{\rm b}$ is almost $q_z$-independent.
Consistently, we recently find that the DW equation eigenvalue for the BO, $\lambda_{\phi}(\q^{\rm 2D}_l,q^z)$, is almost $q_z$-independent by reflecting almost 2D character of 3D 30-orbital CsV$_3$Sb$_5$ model 
\cite{Onari-future}.
(Note that $a_{\rm b}\propto(-1+\lambda_\phi^{-1})$.)
In contrast, the $q_z$-dependence of $b_1$ is rather prominent
\cite{Onari-future}.
Thus, we expect that the namatic 3Q BO state discussed in Ref. \cite{Balents2021} is realistic.
}

\textcolor{black}{
Next, we discuss the 3D structure of the field-induced $3Q$ current order.
In the absence of the BO, by the same argument as above, 
the $3Q$ current order will form the $\pi$-shift stacking 
to gain the $m_2$-term contribution in $\Delta {\bar F}$.
When the $3Q$ current order appears inside the $3Q$ BO state,
the 3D stacking of the current order 
would be mainly determined by the the 3rd order GL term $b_2$
that describes the bond-current coupling energy.
}

\subsection{Derivation of $E_0$}

The expressions of orbital magnetization derived by Refs. \cite{Morb-paper1,Morb-paper2}
is given as
\begin{eqnarray}
&&\!\!\!\!\!\!\!\!
M_{\rm orb}=\frac{ea^2}{2  c \hbar N_{\rm uc}N}\sum_{\k \sigma}m(\k),
\nonumber 
\end{eqnarray}
where $m(\k)$ is given in Eq. [\ref{eqn:Mch-integ}].
$a$ is the unit length in this numerical calculation.
Here, we set $a=|\bm{a}_{\rm AB}|$.
$N_{\rm uc}$ is the site number of the unit cell, 
$N$ is the $\k$-mesh number.
By using the $\mu_B=\frac{e\hbar}{2m_e c}$ and $E_0 \equiv \frac{\hbar^2}{m_e a^2}$,
we obtain Eq. [\ref{eqn:Mch}].
By using the  $m_e c^2=0.511\times10^6$ [eV] and $\frac{\hbar}{m_e c}=3.86 \times 10^{-13}$ [m], we obtain
$E_0=0.5$ [eV] for $a=0.4$ [nm].
In Kagome metals ($a=0.275$ [nm]), $E_0=1.0$ [eV].
In the numerical calculation, large number of $N$ is required to obtain 
a reliable result at low temperatures.
Here, we set $N=(2400)^2$ at $T=1$ [meV].

\subsection{Acknowledgments}
We are grateful to Y. Matsuda, T. Shibauchi, K. Hashimoto, T. Asaba,
S. Onari, A. Ogawa and K. Shimura for fruitful discussions.
This study has been supported by Grants-in-Aid for Scientific
Research from MEXT of Japan (JP20K03858, JP20K22328, JP22K14003),
and by the Quantum Liquid Crystal
No. JP19H05825 KAKENHI on Innovative Areas from JSPS of Japan.


\clearpage
\newpage


\makeatletter
\renewcommand{\thefigure}{S\arabic{figure}}
\renewcommand{\theequation}{S\arabic{equation}}
\makeatother
\setcounter{figure}{0}
\setcounter{equation}{0}
\setcounter{page}{1}
\setcounter{section}{1}

\begin{center}
{\bf \large 
[Supplementary Information] \\
\vspace{3mm}
{\large
Drastic magnetic-field-induced chiral current order  
and emergent current-bond-field interplay
in kagome metal AV$_3$Sb$_5$ (A=Cs,Rb,K)
}
}
\end{center}

\begin{center}
Rina Tazai$^{1*}$, Youichi Yamakawa$^{2*}$, and Hiroshi Kontani$^2$
\end{center}

\begin{center}
\textit{
$^1$Yukawa Institute for Theoretical Physics, Kyoto University,
Kyoto 606-8502, Japan \\
$^2$Department of Physics, Nagoya University,
Nagoya 464-8602, Japan
}
\end{center}


\subsection*{A:
Bandstructure with bond+current orders}\

\textcolor{black}{
In the main text, we study the orbital magnetization $M_{\rm orb}$
in the presence of the current and bond orders
based on Eqs. [\ref{eqn:Mch}] and [\ref{eqn:MchT0}] in the main text
\cite{Morb-paper1S,Morb-paper2S}.
$M_{\rm orb}$ originates from the vertical p-h excitation between
the occupied bands ($\e_{\a\k}<\mu$)
and unoccupied bands ($\e_{\b\k}>\mu$).
In the present study, $M_{\rm orb}$ becomes finite 
in the presence of the TRSB order current order parameters.
Thus, it is important to understand the 
change in the bandstructure due to the order parameters.
Hereafter, the unit of energy is eV unless otherwise noted.
}

\begin{figure}[htb]
\includegraphics[width=.99\linewidth]{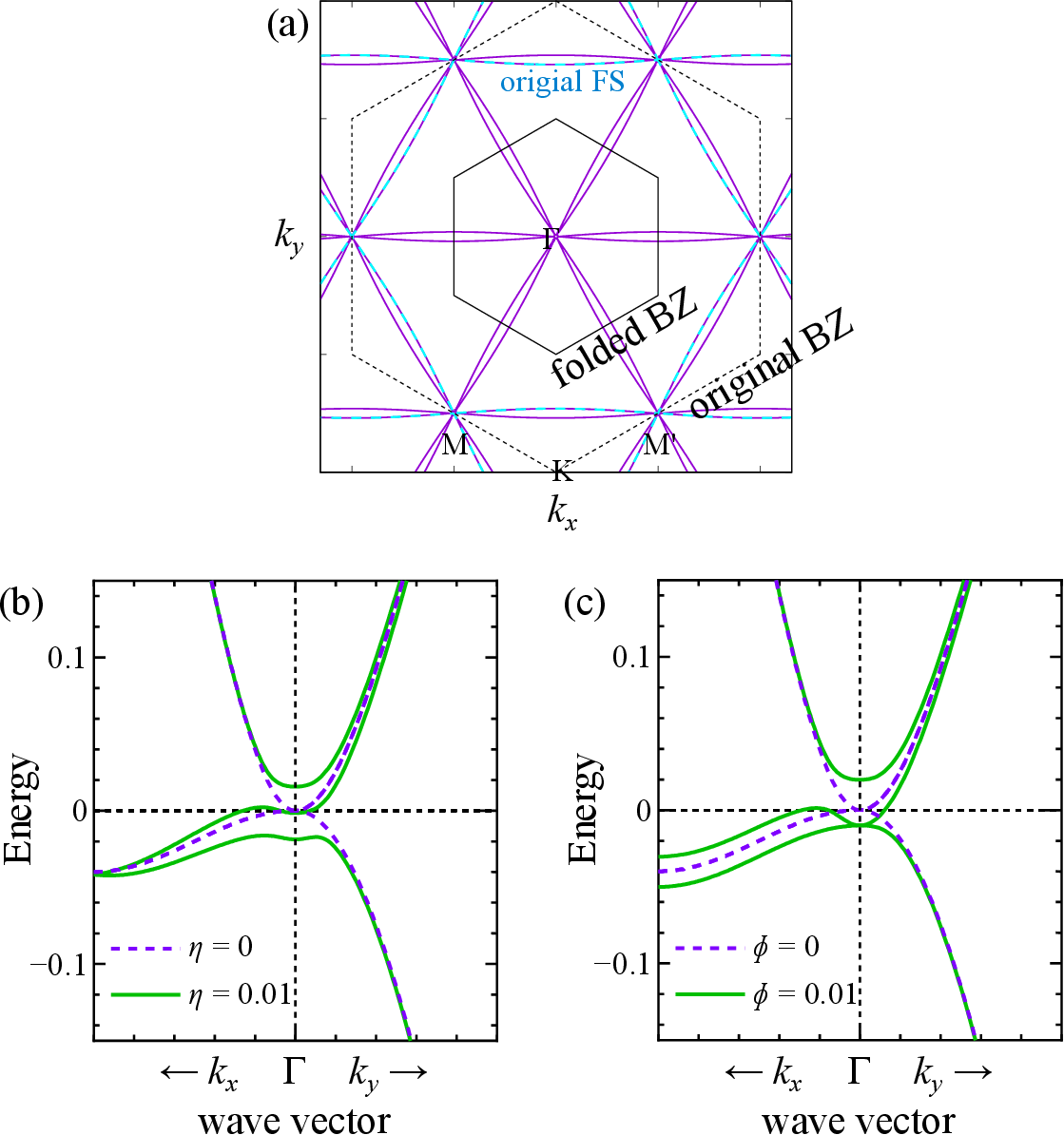}
\caption{
(a) Folded FS at vHS filling for $\phi=\eta=0$.
(b)(c) Folded bandstructure 
in the presence of (b) the $3Q$ current order ($\eta=0$, $0.01$)
and (c) the $3Q$ BO ($\phi=0$, $0.01$).
Here, we set the origin of the energy at the vHS energy.
}
\label{fig:folded-bandS}
\end{figure}

\textcolor{black}{
The folded FS at the vHS filling for $\phi=\eta=0$
is shown in Fig. \ref{fig:folded-bandS} (a).
Figure \ref{fig:folded-bandS} (b) shows
the folded bandstructure at $n=n_{\rm vHS}$
around $\Gamma$ point in the presence of $3Q$ current order
${\bm \eta}=(\eta,\eta,\eta)$
for $\phi=0$ and $0.01$.
Here, all vHS points A, B, C in Fig. \ref{fig:fig1} (b)
move to $\Gamma$ point,
and they split into one bonding ($\delta E=-\sqrt{3}\eta$), 
one antibonding ($\delta E=\sqrt{3}\eta$),
and one unhybridized ($\delta E=0$) states for finite $\eta$.
(The bandstructure for $3Q$ BO ${\bm \phi}=(\phi,\phi,\phi)$
is shown in Fig. \ref{fig:folded-bandS} (c).
The three vHS points split into two bonding ($\delta E=-\phi$) 
and one antibonding ($\delta E=2\phi$).)
}

\textcolor{black}{
In the bandstructure 
shown in Fig. \ref{fig:folded-bandS} (a),
the ``vertical p-h excitations'' (from $\e_{\a\k}<0$ to $\e_{\b\k}>0$ for $\mu=0$)
in the expression of $M_{\rm orb}$
are allowed in a wide $\k$-space for $\mu\sim0$;
around $\Gamma$-M and M-M' lines and around $\Gamma$ and M points.
Then, the factor $(\e_{\a\k}+\e_{\b\k}-2\mu)^{-2}$
in the integrand of $M_{\rm orb}$ is $O(\eta^{-2})$,
and the cancellation due to the factor $(\e_{\a\k}+\e_{\b\k}-2\mu)$
is imperfect due to the p-h asymmetry.
For this reason,
large $|M_{\rm orb}|$ is realized at $n\sim n_{\rm vHS}$.
}

\subsection*{B: 
Relation $m_2\approx-m_3$, Expansion of $M_{\rm orb}[{\eta},{\phi}]$ based on the Green function method}\

\begin{figure}[htb]
\includegraphics[width=.99\linewidth]{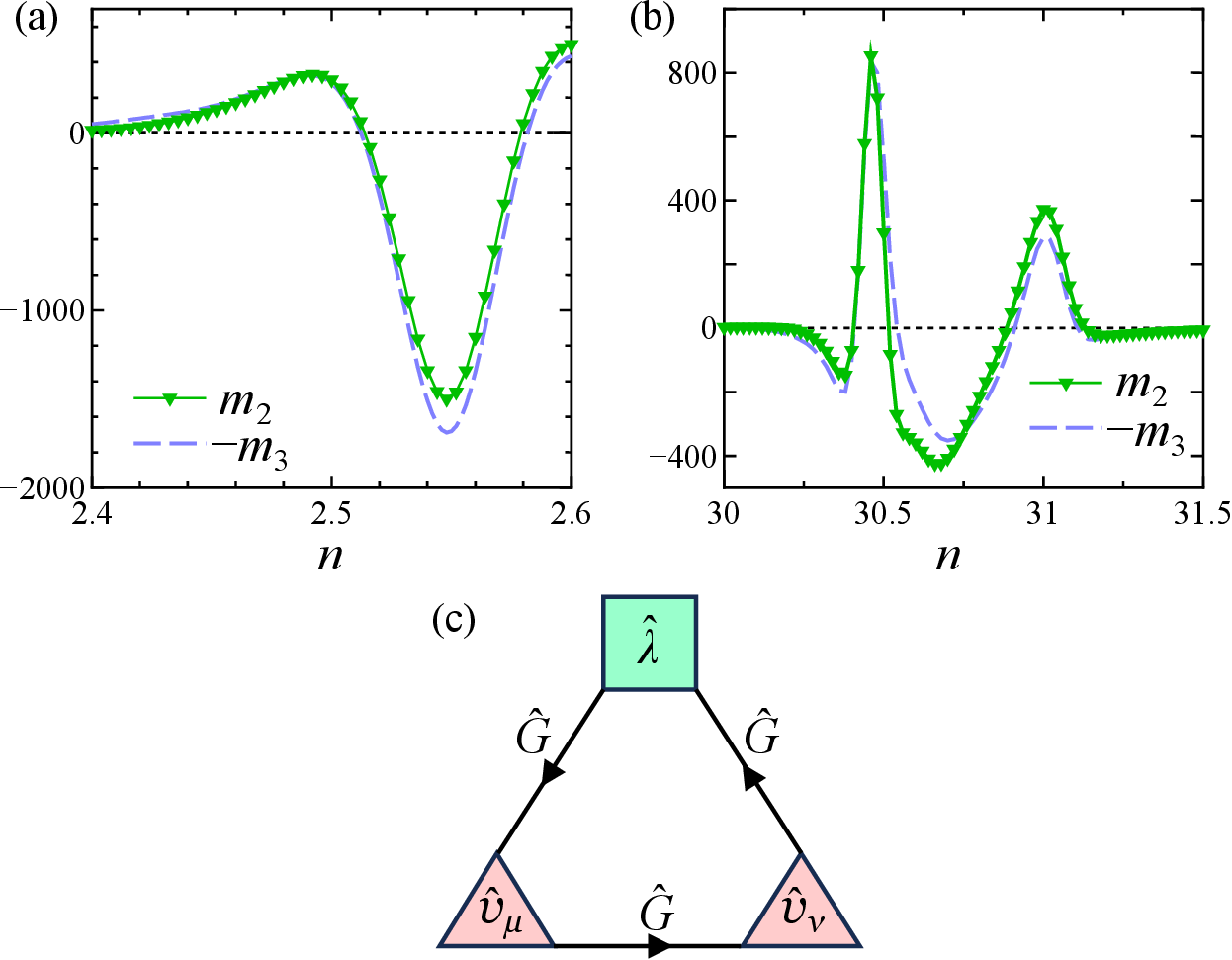}
\caption{
Obtained $m_2$ and $-m_3$ in the 
(a) three-orbital kagome lattice model and 
(b) 30-orbital first-principles model with $\Delta E_p=-0.2$.
(c) Diagrammatic expression of $M_{\rm orb}$ given in Eq. [\ref{eqn:MchT0GS}].
}
\label{fig:m2m3-comp}
\end{figure}

\textcolor{black}{
In Fig. \ref{fig:m2m3-comp}, we show the obtained $m_2$ and $-m_3$ in the 
(a) three-orbital kagome lattice model and 
(b) 30-orbital first-principles model.
In both models, the relation $m_2\approx -m_3$ is well satisfied.
For this reason, we present only $m_1$ and $m_2$ in the main text.
}

\textcolor{black}{
To understand the approximate relation $m_2\approx -m_3$,
we consider the expansion of $M_{\rm orb}$ 
with respect to ${\bm \eta}$ and ${\bm \phi}$;
$M_{\rm orb}=\sum b_{prq}^{p'q'r'}\eta_1^p \eta_2^q \eta_3^r \phi_1^{p'} \phi_2^{q'} \phi_3^{r'}$.
In this notation, $m_2=b^{000}_{111}$ and $m_3=b^{110}_{001}$.
To discuss the nature of the coefficient $b_{prq}^{p'q'r'}$,
we analyze the expression of $M_{\rm orb}$
based on the thermal Green function method
derived in Ref. \cite{KotliarS}:
\begin{eqnarray}
M_{\rm orb}&=&\frac{\mu_{\rm B}}{E_0N_{\rm uc}N} \sum_{ab}^{x,y,z}\e_{zab}
\nonumber \\
& &\times T\sum_{\k,n} {\rm Im}{\rm Tr}
\large( {\hat \lambda}_\k{\hat G}(k) {\hat v}_\k^b 
{\hat G}(k) {\hat v}_\k^c {\hat G}(k) \large) ,
\label{eqn:MchT0GS}
\end{eqnarray}
where ${\hat h}_\k$ is the tight-binding Hamiltonian 
with finite ${\bm \eta},{\bm \phi}$,
${\hat\lambda}_\k={\hat h}_\k-\mu{\hat 1}$, 
$v_\k^a= \d{\hat h}_\k/\d k_a$ is the velocity,
and ${\hat G}(k)$ is the Green function.
$\e_{abc}$ is the antisymmetric tensor, 
and $k\equiv(\k,\e_n)$ with $\e_n=(2n+1)\pi T$.
The diagrammatic expression of Eq. [\ref{eqn:MchT0GS}]
is shown in Fig. \ref{fig:m2m3-comp} (c).
The Hamiltonian of the $4\times3$ site model, 
\begin{eqnarray}
{\hat h}_\k= {\hat h}_\k^0+{\hat \Sigma}_\k,
\label{eqn:HS}
\end{eqnarray}
where ${\hat h}_\k^0$ is the $12\times12$ Hamiltonian for
${\bm\eta}={\bm\phi}={\bm0}$ and 
${\hat \Sigma}_\k^{l,m}= \eta_{l,m}f_\k^{l,m}+\phi_{l,m}g_\k^{l,m}$,
where $f_\k^{l,m}$ [$g_\k^{l,m}$] is the form factor of the current order [BO],
which is given by the Fourier transform of $f_{ij}$ [$g_{ij}$]
introduced in the main text.
$\eta_{l,m}$ [$\phi_{l,m}$] ($l,m=1\sim12$) 
is the current order [BO] parameter.
Then, the Green function is expanded as
\begin{eqnarray}
{\hat G}(k)&=&{\hat G}^0(k)+{\hat G}^0(k){\hat \Sigma}_\k{\hat G}^0(k)
\nonumber \\
& &+{\hat G}^0(k){\hat \Sigma}_\k{\hat G}^0(k){\hat \Sigma}_\k{\hat G}^0(k)
\nonumber \\
& &+\cdots ,
\label{eqn:GS}
\end{eqnarray}
where ${\hat G}^0(k)=((i\e_n-\mu){\hat 1}-{\hat h}^0_\k)^{-1}$.
Thus, the coefficient $m_1$ ($m_2$) can be derived from 
Eq. [\ref{eqn:MchT0GS}], by expanding it with respect to
$\eta_1\phi_1$ ($\eta_1\eta_2\eta_3$)
by using Eqs. [\ref{eqn:HS}]-[\ref{eqn:GS}].
}

\textcolor{black}{
Because of the relations
\begin{eqnarray}
&&h_\k^{0,lm}=h_{-\k}^{0,ml},
\label{eqn:hmkS}
\\
&&G^{0,lm}(\k,\e_m)=G^{0,ml}(-\k,\e_m) ,
\label{eqn:GmkS}
\end{eqnarray}
$M_{\rm orb}$ given by Eq. [\ref{eqn:MchT0GS}] is zero
for ${\bm \phi}={\bm \eta}={\bm0}$.
Equations [\ref{eqn:hmkS}] and [\ref{eqn:GmkS}]
are violated for ${\bm \eta}\ne{\bm0}$
because the current order form factor is odd-parity:
$f_\k^{l,m}=-f_{-\k}^{m,l}=-(f_{-\k}^{l,m})^*$.
(Note that $g_\k^{l,m}=f_{-\k}^{m,l}=(g_{-\k}^{l,m})^*$ for the BO.
Any Hermitian orders with the wavevector $\q={\bm0}$
satisfy $f_\k^{l,m}=(f_{\k}^{m,l})^*$.)
Thus, only odd-order terms with respect to $\eta$ 
can give nonzero $M_{\rm orb}$.
}

\textcolor{black}{
Here, we consider $M_{\rm orb}$ in Eq. [\ref{eqn:MchT0GS}]
in the original 3-site unit cell picture.
Then, the Green function in Eq. [\ref{eqn:MchT0GS}]
is expressed as
${\hat G}(k+q;k)={\hat G}^0(k)\cdot \delta_{\q,{\bm0}}
+\sum_m{\hat G}^0(k+q_m){\hat \Sigma}_k^{(m)}{\hat G}^0(k)\cdot \delta_{\q,\q_m}
+\sum_{m,n}{\hat G}^0(k+q_m+q_n){\hat \Sigma}^{(m)}_{k+q_n}{\hat G}^0(k+q_n){\hat  \Sigma}^{(n)}_k{\hat G}^0(k)\cdot \delta_{\q,\q_m+\q_n} + \cdots$,
where ${\hat G}$ and ${\hat \Sigma}^{(m)}$ are $3\times3$ matrices.
The momentum $\q_m$ is introduced by ${\hat \Sigma}^{(m)}_\k$.
To obtain finite $M_{\rm orb}$,
the total momenta introduced by ${\hat \Sigma}^{(m_i)}$,
$\sum_i^{1,2,\cdots} \q_{m_i}$, should be ${\bm0}$ (modulo original reciprocal vectors).
}

\textcolor{black}{
Hereafter, we discuss the reason for the relation $m_2\approx -m_3$.
In this study, the nearest-neighbor order parameters are
given in Fig. \ref{fig:fig1} (c) for the current order
and Fig. \ref{fig:fig1} (a) for the BO.
In the original 3-site unit cell picture,
the relation $f_{\k'}^{ll'}f_\k^{l'l''}=-g_{\k'}^{ll'}g_\k^{l'l''}$ 
($\k'=\k+(\q_{l'}-\q_{l''})$) holds, where $l\ne l'\ne l''$.
For instance, we consider a $m_2$ term given by replacing
${\hat v}_{x(y)\k}$, ${\hat \lambda}_\k$ and ${\hat G}$
with $(\d/\d k_{x(y)}) {\hat f}_\k$, ${\hat f}_\k$ and ${\hat G}^0$, respectively.
The corresponding $m_3$ term is given by replacing
$(\d/\d k_{x(y)}) {\hat f}_\k$ with $(\d/\d k_{x(y)}) {\hat g}_\k$ in $m_2$.
Due to the absence of sublattice hybridization 
around the vHS points \cite{Thomale2021S},
${\hat G}^0(k)$ is sublattice diagonal around the vHS points,
that is, $G^{0,mm'}(k)=0$ for $m\ne m'$.
Then, the relation $m_2= -m_3$ is satisfied approximately
due to the relation $f_{\k'}^{ll'}f_\k^{l'l''}=-g_{\k'}^{ll'}g_\k^{l'l''}$ 
with $\k'=\k+(\q_{l'}-\q_{l''})$.
}

\textcolor{black}{
We have just started the analysis based on the Green function method.
This is our important future issue.
}

\subsection*{C: 
Analysis of GL free energy with non-analytic $\eta$-linear term}\

In the main text, 
we studied the strong interplay between the bond and current orders
under the magnetic field in kagome metals.
In Fig. \ref{fig:fig4}, we studied the situation 
where bond order transition temperature $T_{\rm b}^0$ 
is higher than the chiral current order one $T_{\rm c}^0$ at $h_z=0$.
We revealed that 
chiral current order emerges at $T=T_{\rm b}^0$ 
under $h_z\sim10^{-4} \ (\sim 1{\rm T})$.
That is, the current-order transition temperature is enlarged to $T_{\rm b}^0$
under small $h_z$, as shown in Fig. \ref{fig:fig3} (b).
The drastic field-induced $3Q$ current order originates from
the non-analytic $\eta$-linear terms in $\Delta {\bar F}$, that is,
$\Delta {\bar F}= -3h_z[m_1{\bm \phi}\cdot{\bm\eta}+m_3(\eta_1\phi_2\phi_3+{\rm cycl.}) + O(\eta^3)]$.

\begin{figure}[htb]
\includegraphics[width=.99\linewidth]{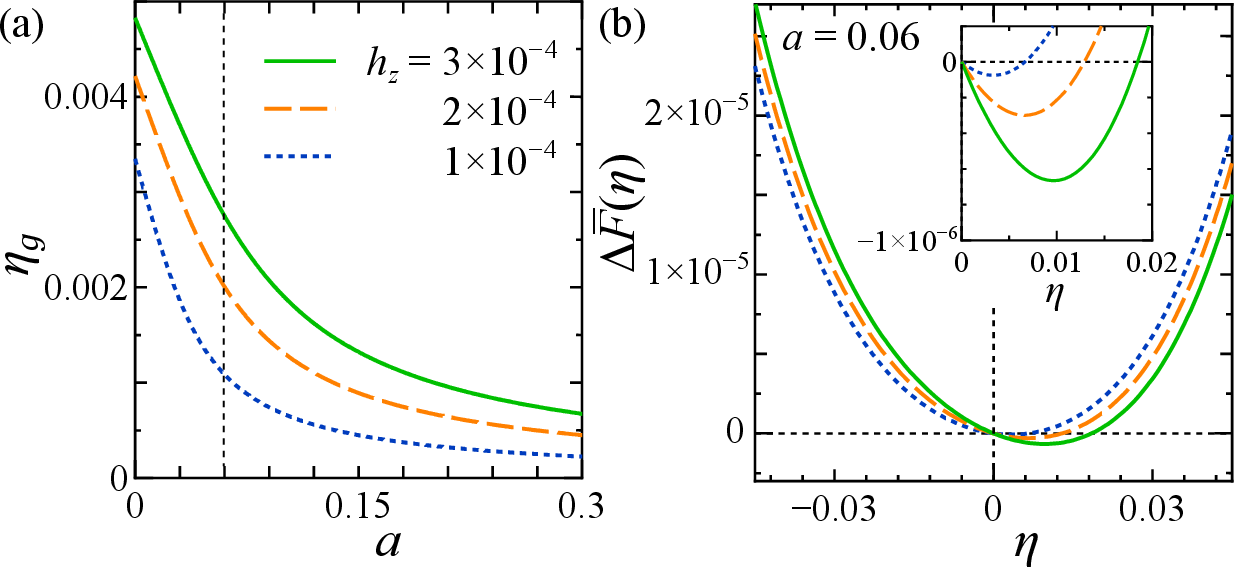}
\caption{
(a) Induced $\eta_g$ due to the $\eta$-linear term for as a function of $a \ (>0)$.
Here, $g=0.9h_z$ and $h_z=1\sim3\times10^{-4}$.
This result is not sensitive to the choice of $d$ when $g$ is small.
(b) $\Delta {\bar F}$ as a function of $\eta$ for $a=0.06$.
(Inset) Enlarged plots around the bottoms.
}
\label{fig:figS3}
\end{figure}

To understand the effect of the field-induced 
non-analytic free energy qualitatively,
we analyze the following simple GL free energy with a $\eta$-linear term:
\begin{eqnarray}
F(\eta)= a\eta^2+d\eta^4-g\eta .
\label{eqn:simpleGL}
\end{eqnarray}
Here, we assume $a$ and $d$ are positive.
When $g\ne0$, $F(\eta)$ is minimized at finite $\eta$ even if $a>0$.
The solution is given as
\begin{eqnarray}
\eta_{g}&=&X^{1/3}(d^{-1}-2\cdot 3^{1/3}a X^{-2/3})/(2\cdot 3^{2/3}),
\\
X&=&9gd^2+\sqrt{3}\sqrt{d^3(8a^3+27dg^2)} .
\label{eqn:eta-min}
\end{eqnarray}
Note that 
\begin{eqnarray}
\eta_g\approx g/2a ,
\end{eqnarray}
for small $g$.
Therefore, $\eta_g$ is finite even when $a>0$
due to the $\eta$-linear term.

Here, we derive $d$ and $g$ from the GL free energy of kagome metal,
Eqs. [\ref{eqn:Feta}] and [\ref{eqn:Fphi}] in the main text.
By setting ${\bm \phi}=(\phi,\phi,\phi)$ and 
${\bm \eta}=(\eta,\eta,\eta)$ in Eqs. [\ref{eqn:Feta}] and [\ref{eqn:Fphi}],
we obtain 
$d=6d_{i}$ and $g=3(3m_1\phi+3m_3\phi^2)h_z$.
When $d_i=150$, $m_1=5$, $m_3=1000$ and $\phi=0.01$, 
we get $d=900$ and $g=1.35h_z$.
$a$ in Eq. [\ref{eqn:simpleGL}]
corresponds to $3a_{\rm c}$ in the main text.
In Fig. \ref{fig:fig3} (b) in the main text,
we set $a_{\rm c}=r_{\rm c}(T-T_{\rm c}^0)$ with $r_{\rm c}=30$ and $T_{\rm c}^0=0.005$,
so $T=T_{\rm c}^0 \ (2T_{\rm c}^0=T_{\rm b}^0)$ in Fig. \ref{fig:fig4} (b)
corresponds to $a=0 \ (0.45)$ in Eq. [\ref{eqn:simpleGL}].


Figure \ref{fig:figS3} (a) shows $\eta_g$ given in Eq. [\ref{eqn:eta-min}]
as a function of $a$, in the case of $d=1000$ and $g=1.35h_z$.
We see that $h_z\sim 10^{-4}$, which corresponds to $\sim1$T,
induces sizable current order even above $T_{\rm c}^0$ if $a\gtrsim0$.
The obtained value of $\eta_g$ in Fig \ref{fig:figS3} (a) at $a\approx0$ 
is comparable to the field-induced order in Fig. \ref{fig:fig4} (b).
The field-induced $\eta$ is prominent 
only when the system at $h_z=0$ is close to the current order state
({\it i.e.}, $a\gtrsim0$).
Asymmetric $\Delta {\bar F}$ as a function of $\eta$ is shown in 
Fig \ref{fig:figS3} (b).

\subsection*{D: 
Comparison between ${\bar M}_{\rm orb}$ and $M_{\rm orb}$}\

In the main text, we calculated the orbital magnetization 
$M_{\rm orb}$ in kagome metal with current order ${\bm \eta}$
and bond-order ${\bm \phi}$ using Eq. [\ref{eqn:Mch}].
Next, we derived its expression 
up to the third-order of ${\bm \eta}$ and ${\bm \phi}$:
${\bar M}_{\rm orb}= m_1 {\bm \phi}\cdot{\bm \eta} + m_2 \eta_1\eta_2\eta_3
+m_3(\eta_1\phi_2\phi_3 +({\rm cycl.}))$.
\textcolor{black}{
To obtained the coefficients $m_1$ and $m_2$,
we calculate $M_{\rm orb}[{\bm \eta},{\bm \phi}]$ very accurately
and expand it around ${\bm \eta}={\bm \phi}=0$ numerically.
For instance, we derive $m_2$ as
$m_2= M_{\rm orb}[(\eta_1,\eta_2,\eta_3),{\bm0}]/\eta_1\eta_2\eta_3$
with $\eta_m=0.001$.
(Note that $M_{\rm orb}[{\bm\eta},{\bm0}]=0$ if one of $\eta_m$ is zero.)
}

The expression ${\bar M}_{\rm orb}$ is very useful to understand 
the strong interplay between current and bond orders in kagome metal.
By considering the field-induced free energy 
$\Delta {\bar F}=-3h_z {\bar M}_{\rm orb}$,
we understand the characteristic phase diagram of kagome metals
under the magnetic field.

\begin{figure}[htb]
\includegraphics[width=.99\linewidth]{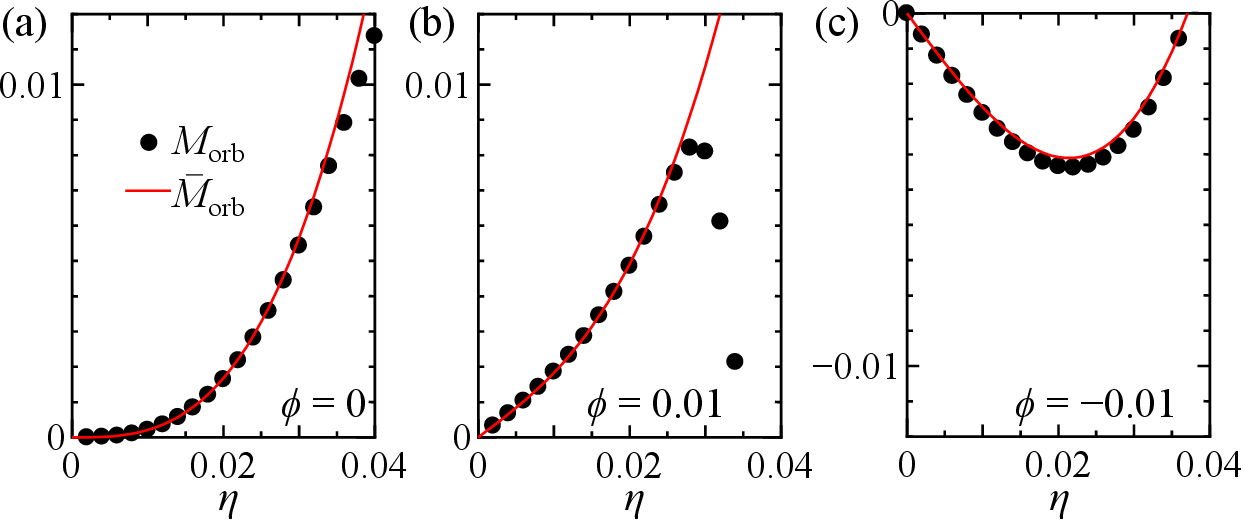}
\caption{
$M_{\rm orb}$ and ${\bar M}_{\rm orb}$ for
${\bm\eta}=(\eta,\eta,\eta)/\sqrt{3}$
in the cases of (a) ${\bm\phi}={\bm0}$ and 
(b) ${\bm\phi}=(\phi,\phi,\phi)/\sqrt{3}$ at $\phi=+0.01$ and (c) $-0.01$.
}
\label{fig:figS1}
\end{figure}

Here, we verify that ${\bar M}_{\rm orb}$ in Eq. [\ref{eqn:M-exp}]
well reproduces $M_{\rm orb}$ in Eq. [\ref{eqn:Mch}].
Figures \ref{fig:figS1} (a)-(c) show the obtained results at $n=2.47$
under ${\bm \eta}=(\eta,\eta,\eta)/\sqrt{3}$ and 
${\bm \phi}=(\phi,\phi,\phi)/\sqrt{3}$
as functions of $\eta$,
in the case of (a) $\phi=0$, (b) $\phi=+0.01$ and (b) $\phi=-0.01$.
It is found that ${\bar M}_{\rm orb}$
well reproduce the original $M_{\rm orb}$ 
when $|{\bm \eta}|,|{\bm \phi}|\lesssim 0.02$,
unless the shape of the Fermi surface is drastically
changed by order parameters.

\begin{figure}[htb]
\includegraphics[width=.99\linewidth]{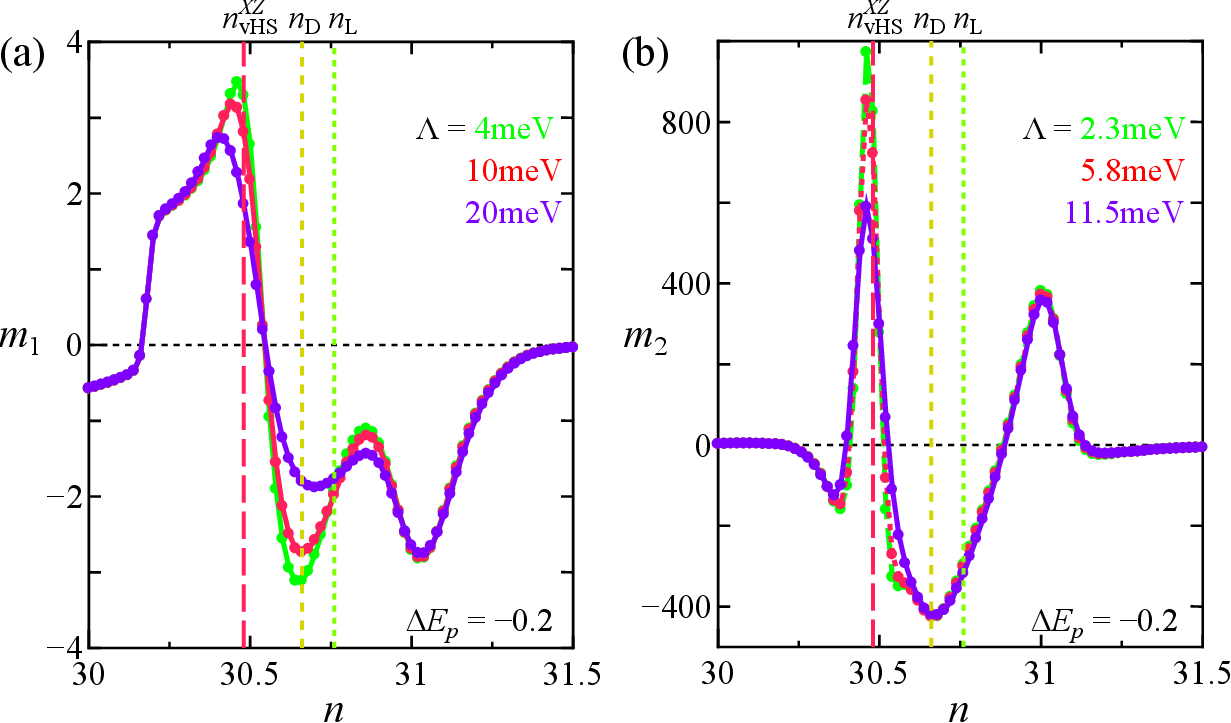}
\caption{
Obtained coefficients
(a) $m_1$ and (b) $m_2$ at $T=0.005$
derived from the region $|\eta_i|,|\phi_i|<\Lambda$.
}
\label{fig:m1m2-30orbitalS}
\end{figure}

\textcolor{black}{ 
Next, we examine the validity of the 
expansion expression ${\bar M}_{\rm orb}$
in the realistic 30-orbital model for kagome metal. 
Figure \ref{fig:m1m2-30orbitalS} shows the coefficients 
$m_1$ and $m_2$ given by $M_{\rm orb}$ in Eq. [\ref{eqn:Mch}]
in the main text, derived from the region $|\eta_i|,|\phi_i|<\Lambda$.
The convergence of the obtained results 
for both $m_1$ and $m_2$ is good for $\Lambda<0.01$
for wide range of $n$, except at the close vicinity of the vHS filling.
Therefore, the GL free energy expression [\ref{eqn:DeltaF2}]
in the main text is valid for real kagome metals.
}

\subsection*{E: 
$M_{\rm orb}$ in kagome lattice model with $(t,t')=(-0.5,-0.08)$}\

In the main text, we studied $M_{\rm orb}$ 
in kagome lattice model with the bare hopping integrals 
$t=-0.5$ and $t'=-0.02$.
The obtained coefficients $m_1$ and $m_2$ in ${\bar M}_{\rm orb}$
take large values for $n\approx n_{\rm vHS}$, where $n_{\rm vHS}$ is the van-Hove filling.
Using the obtained coefficients $m_1$ and $m_2$,
we discovered the mechanism of the field-induced chiral current order in kagome metals.

\begin{figure}[htb]
\includegraphics[width=.99\linewidth]{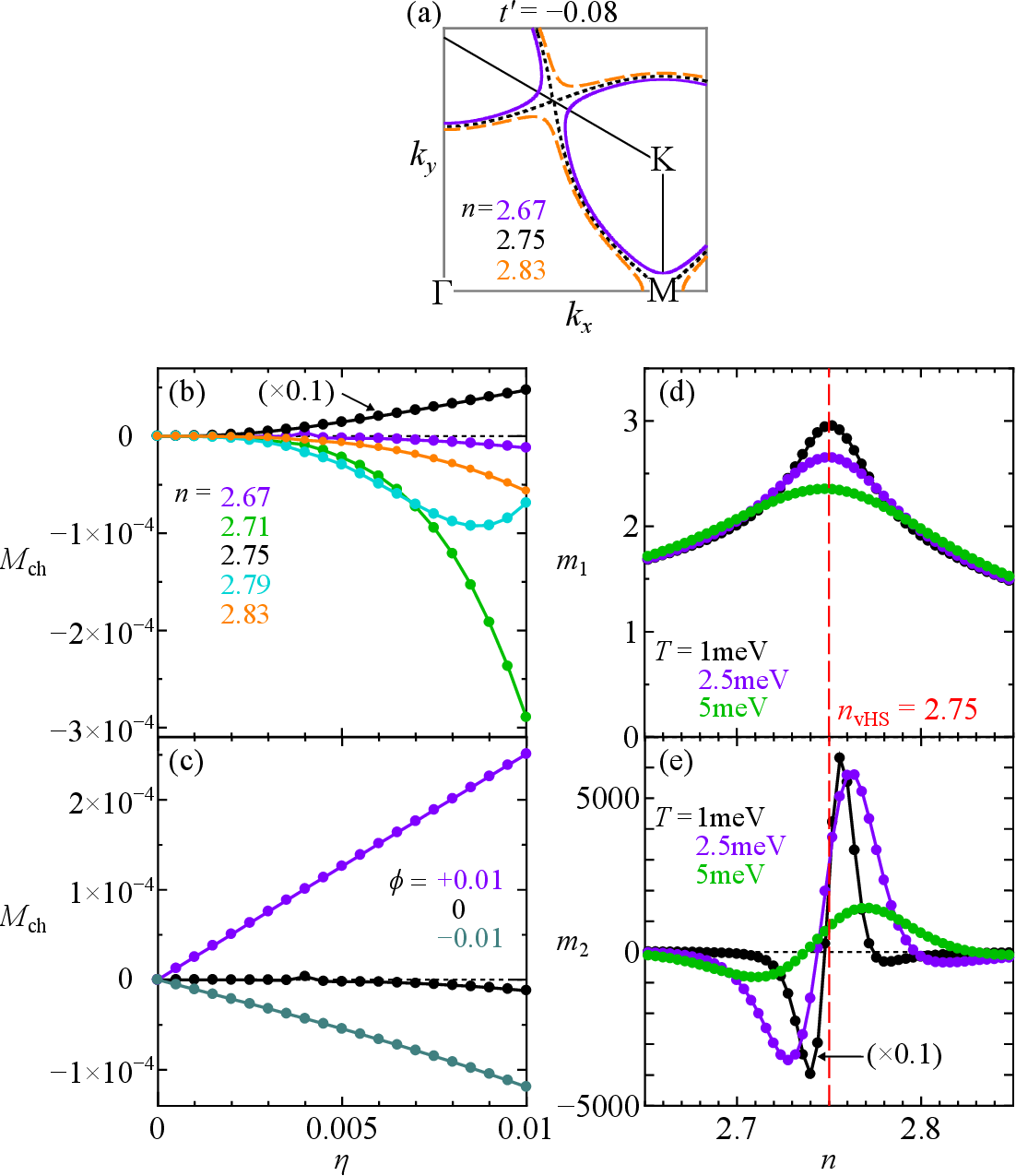}
\caption{
(a) FS around $n_{\rm vHS}=2.75$ at $t'=-0.08$.
(b) $M_{\rm orb}$ [$\mu_{\rm B}$] per $V$ atom due to the
$3Q$ current order ${\bm \eta}=(\eta,\eta,\eta)/\sqrt{3}$ 
at $T=1$ meV for $n=2.67\sim2.83$.
(c) $M_{\rm orb}$ due to the coexistence of
$3Q$ current order and $3Q$ BO at $n=2.67$.
Obtained (d) $m_1$ and (e) $m_2\ (\approx-m_3)$,
per $V$ atom as a function of $n$.
Both $|m_1|$ and $|m_2|$ are large for $n\sim n_{\rm vHS}$.
}
\label{fig:figS2}
\end{figure}

To verify the robustness of the obtained results,
here we analyze $M_{\rm orb}$ 
in kagome lattice model with large $t'$; $(t,t')=(-0.5,-0.08)$.
The obtained FS shown in Fig. \ref{fig:figS2} (a)
has large curvature due to large $|t'|$.
Here, we have introduce the $3Q$ BO ${\bm \phi}=(\phi,\phi,\phi)/\sqrt{3}$
and the $3Q$ current order ${\bm \eta}=(\eta,\eta,\eta)/\sqrt{3}$.
Figure \ref{fig:figS2} (b) shows the obtained  $M_{\rm orb}$ [$\mu_{\rm B}$] 
per $V$ atom due to the $3Q$ current order with $\phi=0$.
The relation $M_{\rm orb}\propto \eta^3$ is satisfied,
and its magnitude is enlarged when $n\approx n_{\rm vHS}$.
Figure \ref{fig:figS2} (c) shows $M_{\rm orb}$ [$\mu_{\rm B}$] 
due to the coexistence of $3Q$ current order and $3Q$ BO.
The relation $M_{\rm orb}\propto \eta^1$ is satisfied when $\phi\ne0$.
Figures \ref{fig:figS2} (d) and (e) represent the obtained 
coefficients $m_1$ and $m_2$ in ${\bar M}_{\rm orb}$.
The magnitudes of $m_1$ and $m_2$ for $t'=-0.08$
are comparable to those for $t'=-0.02$ given in the main text.

\subsection*{F: 
Calculation of GL parameters, Renormalization of $a_{\rm c}$}\

\begin{figure}[htb]
\includegraphics[width=.99\linewidth]{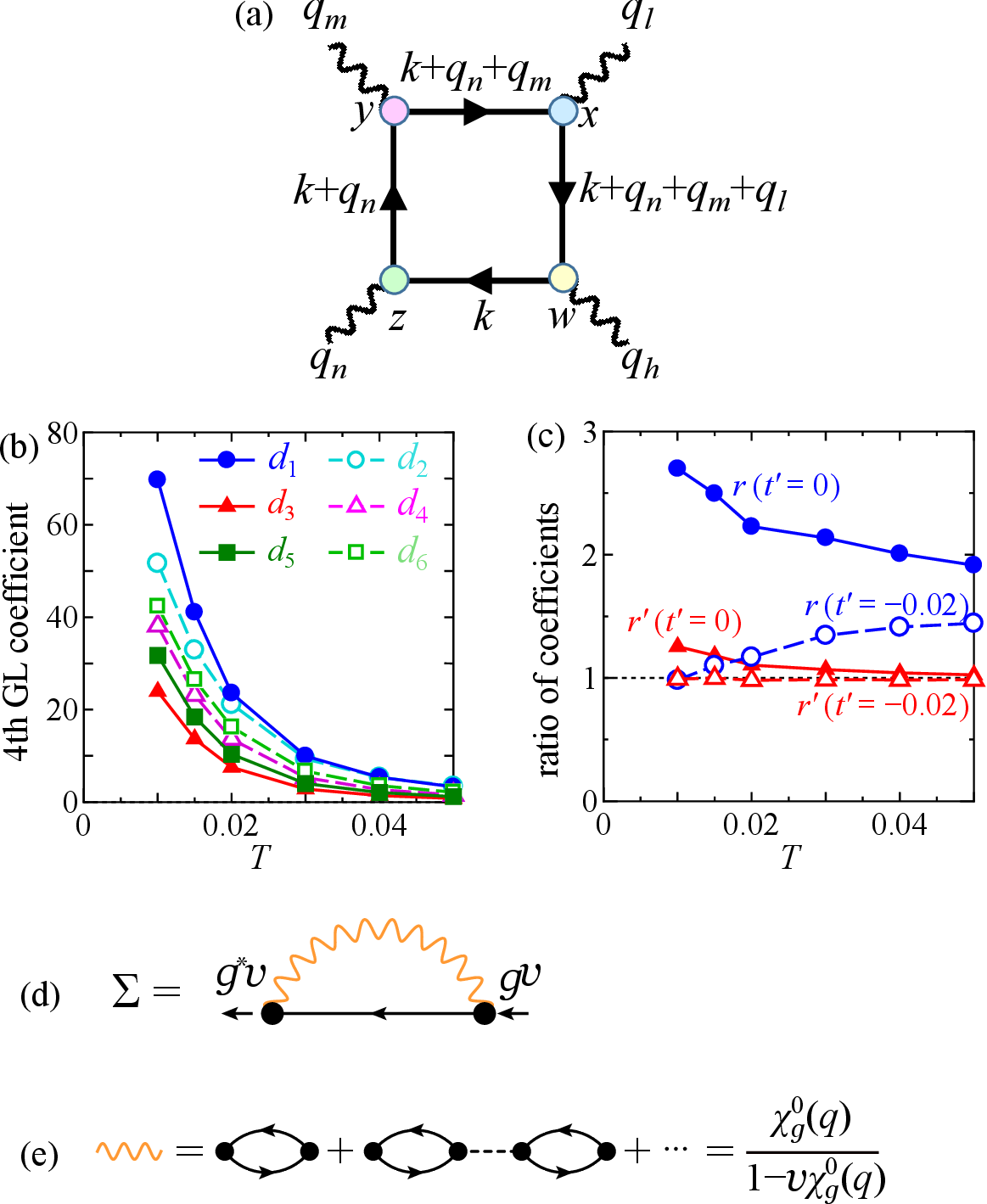}
\caption{
(a) Diagrammatic expressions of 4th order GL parameters.
(b) Numerical results of the 4th order GL parameters as functions of $T$.
(c) Ratio $r=2d_1/d_2$ and $r'=2d_3/d_4$ in the case of $t'=0$ and $t'=-0.02$.
(d) Diagrammatic expression of the self-energy driven by the 
BO fluctuations.
(e) Expression of BO susceptibility.
}
\label{fig:figS5}
\end{figure}

\textcolor{black}{
To verify that the GL free energy coefficients
assumed in the main text are qualitatively reasonable,
we calculate GL coefficients based on the diagrammatic method.
The 4rh order GL parameters per unit cell are given as
\begin{eqnarray}
d_{1}&=& I_{1111}^{g},
\\
d_{2}&=& 2I_{1212}^{g}+4I_{1122}^{g},
\\
d_{3}&=& I_{1111}^{f},
\\
d_{4}&=& 2I_{1212}^{f}+4I_{1122}^{f},
\end{eqnarray}
where
\begin{eqnarray}
I_{hlmn}^{w}&=&\frac{T}{4}\sum_{k,\s} {\rm Tr}
{\hat w}_{\q_h}(k+\q_n+\q_m+\q_l)
\nonumber \\
&\times&
{\hat G}(k+\q_n+\q_m+\q_l)
\nonumber \\
&\times&{\hat w}_{\q_l}(k+\q_n+\q_m){\hat G}(k+\q_n+\q_m)
\nonumber \\
&\times&{\hat w}_{\q_m}(k+\q_n)
{\hat G}(k+\q_n){\hat w}_{\q_n}(k){\hat G}(k) ,
\label{eqn:I4}
\end{eqnarray}
where ${\hat G}(k)$ is $3\times3$ Green function 
for the original 3-site kagome lattice model and $k\equiv(\k,i\e_n)$.
The diagrammatic expression $I_{hlmn}^{w}$ 
is depicted in Fig. \ref{fig:figS5} (a).
${\hat w}_\q(\k)$ is the form factor 
in the momentum space in the original BZ,
${\hat f}_\q(\k)$ or ${\hat g}_\q(\k)$, and $h,l,m,n$ is 1, 2, or 3.
Note that $f^{lm}_{\q_m}(\k)$ [$f^{lm}_{\q_m}(\k)$]
($l,m$=1-3) is given by the Fourier transform
of $f_{i,j}^{(m)}$ [$g_{i,j}^{(m)}$] introduced in the main text.
The relation $\q_h+\q_l+\q_m+\q_n=0$ should be satisfied.
Hereafter, we use ${\hat f}_\q(\k)$ derived from the 
density-wave equation in Ref. \cite{Tazai-kagome2S},
in which distant-atom components are included.
On the other and, ${\hat g}_\q(\k)$ is derived from
the nearest-neighbor BO in Fig. \ref{fig:fig1} (a).
In the same way, the analytic expressions of $d_5$ and $d_6$ 
are obtained in Ref. \cite{Tazai-kagome2S}.
}

\textcolor{black}{
The obtained numerical results are given in Fig. \ref{fig:figS5} (b)
for $t'=0$ and $t'=-0.02$.
Here, the dimensionless form factors are normalized as
$\max_{\k}|g_{\q_1}^{\rm BA}(\k)| = |g_{\q_1}^{\rm BA}(\k_{\rm A})| =1$.
(This normalization corresponds to $|\delta t_{ij}^{\rm b}|=1/2$.)
In the same way, we set
$\max_{\k}|f_{\q_1}^{\rm BA}(\k)| = |f_{\q_1}^{\rm BA}(\k_{\rm A})| =1$.
Thus, the parameter $d_i=150 \ [{\rm eV^{-3}}]$ is consistent with 
Fig. \ref{fig:figS5} (b) for $T\sim0.01$.
For ${\bm\phi}=(\phi,\phi,\phi)$ with $\phi=0.01$, 
the 4th order term for the BO is
$F_4^{\rm b}\sim 150\cdot 6(10^{-2})^4\sim10^{-5}$ [eV] for $d_i=150$.
The obtained ratios $r=2d_1/d_2$ and $r'=2d_3/d_4$ are shown in 
Fig. \ref{fig:figS5} (c):
For $t'=0$, we obtain $2d_1/d_2\gg1$ and $2d_3/d_4\gtrsim1$. 
For $t'=-0.02$, $2d_1/d_2\sim1$ and $2d_3/d_4\approx1$. 
Both ratios tend to become smaller than unity for larger $|t'|$.
(Note that $d_1\approx d_3 \approx d_5$ and $d_2 \approx d_4 \approx d_6$
when the form factors satisfy $f_{ij}^{(m)}=\pm ig_{ij}^{(m)}$.)
}

\textcolor{black}{
In deriving $d_1\sim d_6$ in Fig. \ref{fig:figS5} (b),
we included the self-energy due to the BO fluctuations
\cite{Tazai-kagome2S},
because the self-energy reduces unrealistic behaviors of $d_i$
at low temperatures when the inter-sublattice nesting vector 
is not exactly commensurate
\cite{Balents2021S}.
(Note that the present method works well only for $|t'/t|\ll1$.)
To calculate the self-energy,
we introduce the following effective BO interaction
\cite{Tazai-kagome2S}:
\begin{eqnarray}
\hat{H}_{\rm int} &=& -\frac{1}{N} \sum_{\q} \frac{v}{2} \,\, \hat{O}^g_{\q} \,\, \hat{O}^g_{-\q} , 
\label{eqn:elph1} 
\end{eqnarray}
where 
$\hat{O}^g_{\q} \equiv \sum_{\k,l,m,\sigma} g_{\q}^{lm} (\k) c^{\dagger}_{\k+\q,l, \sigma} c_{\k,m,\sigma}$
is the BO operator \cite{Tazai-rev2021S,Kontani-AdvPhysS,Kontani-RPAS}.
and $v$ is the effective interaction.
Here, the BO form factor $g_{\q}^{lm} (\k)$ is 
normalized as $\max_{\k,l,m} |g_{\q}^{lm} (\k)|=1$ at each $\q$,
{\i.e.}, $|\delta t_{ij}^{\rm b}|\equiv1/2$ for the nearest sites.
We next calculate the on-site self-energy due to BO fluctuations as
\cite{Tazai-kagome2S}
\begin{eqnarray}
\Sigma_{m}(\e_n)&=&\frac{T}{N^2}\sum_{\k,q,m'',m'''}G_{m'm''}(\k+\q,\e_n+\w_l) 
\nonumber \\
& &\times B_{mm',mm''}(k,q)
\label{eqn:Self}, \\
B_{ml,m'l'}(k,q)&=& g_\q^{lm}(\k)^* g_{\q}^{l'm'}(\k)\cdot 
v(1+v\chi_g(q)) ,
\end{eqnarray}
which is shown in Fig. \ref{fig:figS5} (d).
Here, the BO susceptibility $\chi_g(q)$ is 
\cite{Tazai-kagome2S}
\begin{eqnarray}
\chi_g (q) 
&=& \chi^0_{g}(q)/(1-v \chi^0_{g}(q)),
\label{eqn:chic},
\\
\chi^{0}_g (q) 
&=& \sum_{lmm'l'}  \chi^{0, lmm'l'}_{g} (q) ,
\label{eqn:chi0} \\
\chi^{0,lmm'l'}_g (\q,\omega_l) 
&=& \frac TN \sum_{\k,\e_n} g^{lm}_{\q} (\k)^* G_{lm'}(\k+\q,\e_n+\omega_l)
 \nonumber \\
&& \times  G_{l'm}(\k,\e_n) g^{m'l'}_{\q} (\k) ,
\label{eqn:chi0_2}
\end{eqnarray}
which is shown in Fig. \ref{fig:figS5} (e).
Then, the Green function is given as
${\hat G}(k)=(i\e_n+\mu-{\hat h}(\k)-{\hat \Sigma}(\e_n))^{-1}$.
The effect of thermal fluctuations described by the self-energy 
is essential to reproduce the 
$T$-dependence of various physical quantities.
In the present numerical study,
we calculate Eqs. [\ref{eqn:Self}]-[\ref{eqn:chi0_2}] self-consistently.
}

\textcolor{black}{
We also study the 2nd order GL term, which is derived as
$F_2^{\rm b}\sim -\chi_g^0(\q_1)R|{\bm\phi}|^2$
according to Ref. \cite{Tazai-LWS}.
Here, $\chi_g^0(\q_1)$ is the irreducible BO susceptibility,
and $R\equiv (d\lambda/dT)\cdot (-T_{\rm b}^0)$, where
$\lambda$ is the eigenvalue of the density-wave equation,
which is similar to the eigenvalue of the BCS gap equation.
$R\sim O(0.1)$ in usual BO phase transitions
 \cite{Tazai-LWS},
while $R\sim1$ in BCS superconductivity
because of large ${\rm log}T$ singularity of the p-p channel.
As a result, $F_2^{\rm b}\sim - 3\times 10^{-5} \ [{\rm eV}]$
for $\chi_g^0(\q_1) \sim3$ and $R=0.1$. 
Note that $\chi_g^0(\q_1)R\sim0.1$ corresponds to 
$r T_{\rm b}^0\sim0.1$ in the main text.
}

\textcolor{black}{
Therefore, the BO total free energy is
$F_{\rm tot}^{\rm b}\sim -10^{-5} \ [{\rm eV}]$ for 
$T_{\rm b}^0\sim\phi\sim0.01$ and $N(0)\sim1$.
The current-order total free energy
$F_{\rm tot}^{\rm c}$ will be comparable to $F_{\rm tot}^{\rm b}$.
Thus, the GL parameters assumed in the main text
are qualitatively reasonable.
(In BCS superconductors, 
$F_{\rm tot}= -\Delta^2N(0)/2\sim -10^{-4}$ 
when $\Delta=0.01$ and $N(0)\sim1$.)
As discussed in Ref. \cite{Tazai-LWS},
the specific heat jump $\Delta C/T$ 
at the BO or current-order phase transition is 
much smaller than the BCS value $\Delta C/T_{\rm SC}=1.43N(0)$.
}

\textcolor{black}{
Finally, we discuss the renormalization of the
2nd order GL coefficient for $\eta$, $a_{\rm c}$, in the BO phase.
Under the $3Q$ BO phase ${\bm\phi}=(\phi,\phi,\phi)/\sqrt{3}$,
$a_{\rm c}$ in Eq. [\ref{eqn:Feta}] is renormalized 
as ${\bar a}_{\rm c}=a_{\rm c}+(d_5+d_6)(2\phi^2/3)$
due to the $d_5$, $d_6$ terms.
When ${\bm\eta}={\bm0}$, we obtain $\phi^2\approx-3a_b/2(d_1+d_2)$ by neglecting $b_1$ term, which is allowed except for $T\approx T_{\rm b}^0$.
Therefore, we obtain
${\bar a}_{\rm c}={\bar r}_{\rm c}(T-{\bar T}_{\rm c}^0)$,
where ${\bar r}_{\rm c}=r_{\rm c}(1-C)$,
${\bar T}_{\rm c}^0=(T_{\rm c}^0-CT_{\rm b}^0)/(1-C)$,
and $C\equiv (r_{\rm b}/r_{\rm c})[(d_5+d_6)/(d_1+d_2)]$.
Here, we assume $C<1$ by referring to the relation 
$d_5+d_6<d_1+d_2$ in Fig. \ref{fig:figS5},
owing to the difference between the BO and current order form factors.
For detail, see Ref. \cite{Tazai-kagome2}.
}

\subsection*{G: 
$M_{\rm orb}$ by intra-original-unit-cell current order in kagome lattice}\

\begin{figure}[htb]
\includegraphics[width=.99\linewidth]{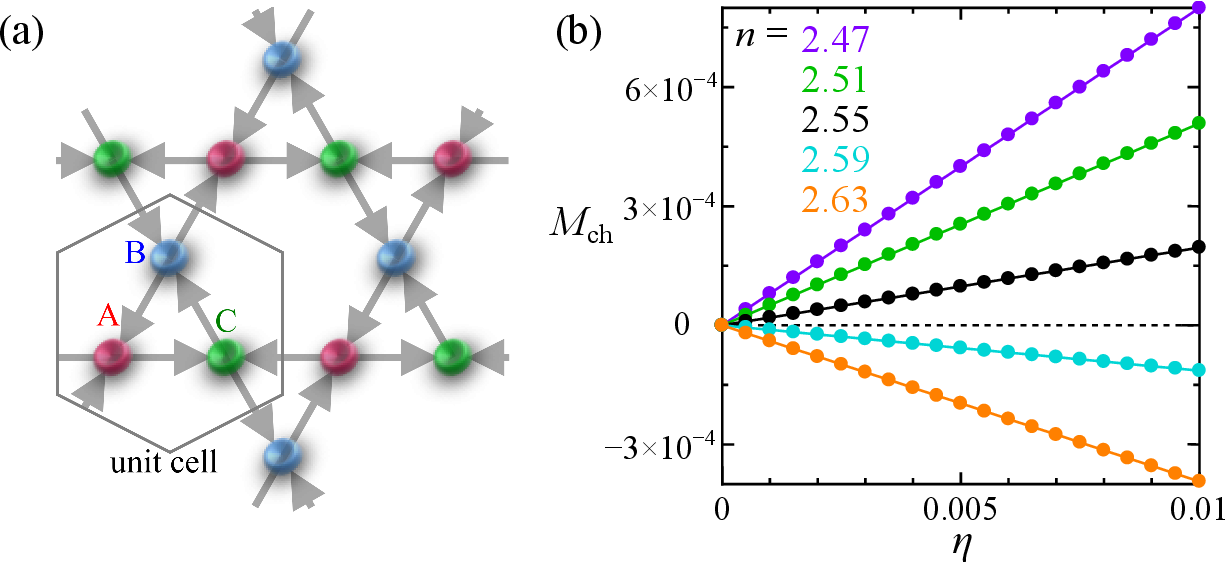}
\caption{
(a) Intra-original-unit-cell ($\q={\bm0}$) current order.
(b) $M_{\rm orb}$ for the intra-original-unit-cell current order
${\bm\eta}=(\eta,\eta,\eta)/\sqrt{3}$.
We set $(t,t')=(-0.5,-0.02)$, and no BO is introduced.
}
\label{fig:figS4}
\end{figure}

Here, we calculate $M_{\rm orb}$ in the case of the 
intra-original-unit-cell ($\q={\bm0}$) 
current order in Fig. \ref{fig:figS4} (a).
In this case, the translational symmetry is preserved.
The obtained $M_{\rm orb}$ is shown in Fig. \ref{fig:figS4} (b).
We find that
$M_{\rm orb}$ is $\eta$-linear even in the absence of the BO,
while its coefficient is small for $n\sim n_{\rm vHS}$
that is realized in kagome metals.
Regardless of the presence of the $\eta$-linear term in $M_{\rm orb}$,
field-induced intra-original-unit-cell order will be quite small in kagome metals.
In fact, 
the field-induced cLC order at $\q={\bm0}$ becomes sizable only when
its second-order GL coefficient,
$a_{\rm c}\sim (-1+\lambda_{\q={\bm0}}^{-1})$, is very small.
Here, $\lambda_{\q={\bm0}}$ is the eigenvalue of the current order 
solution at $\q={\bm0}$.
However, the relation $\lambda_{\q={\bm0}}\ll1$ is obtained 
in the DW equation analysis in Ref. \cite{Tazai-kagome2S}.

\subsection*{H: 
First principles $30$-orbital model for kagome metal}\

\begin{figure}[htb]
\includegraphics[width=.99\linewidth]{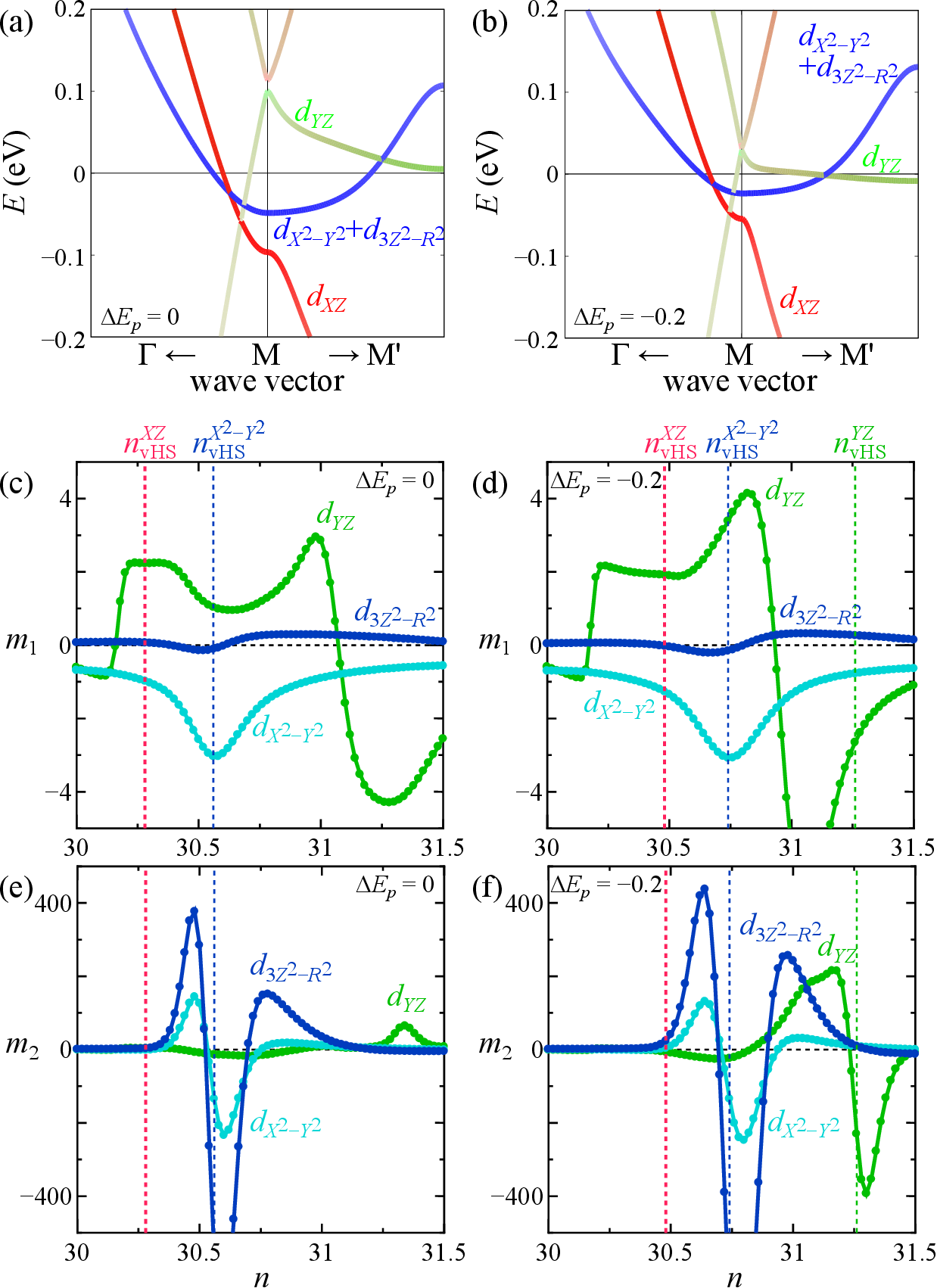}
\caption{
(a) Bandstructure of CsV$_3$Sb$_5$ model in the $k_z=0$ plane.
Its FS is shown in Fig. \ref{fig:fig6} (a).
The $d_{XZ}$-orbital band gives the pure-type FS,
and the $d_{YZ}$-orbital band gives the mix-type FS.
The $d_{X^2-Y^2}$+$d_{3Z^2-R^2}$-orbital band gives another pure-type FS.
$\Gamma$, M, M' are introduced in Fig. \ref{fig:fig1} (b) in the main text.
(b) Bandstructure with the $p$-orbital shift $\Delta E_{p}=-0.2$ eV.
Its FS is shown in Fig. \ref{fig:fig6} (a).
(c)(d) Obtained $m_1$ for the order parameters in the 
$d_{YZ}$, $d_{X^2-Y^2}$, and $d_{3Z^2-R^2}$ orbitals
for (c) $\Delta E_{p}=0$ model and (d) $\Delta E_{p}=-0.2$ model.
(e)(f) Obtained $m_2$ for the order parameters in the 
$d_{YZ}$, $d_{X^2-Y^2}$, and $d_{3Z^2-R^2}$ orbitals
for (e) $\Delta E_{p}=0$ model and (f) $\Delta E_{p}=-0.2$ model.
In (c)-(f), the vHS filling $n_{\rm vHS}^\Gamma$ 
($\Gamma=XZ,YZ,X^2-Y^2$) is shown.
(Note that $n_{\rm vHS}^{X^2-Y^2}=n_{\rm vHS}^{3Z^2-R^2}$.)
}
\label{fig:m1m2S}
\end{figure}

\textcolor{black}{
In the main text, we analyzed the GL coefficients 
based on the first-principles 30-orbital model for kagome metals.
Figure \ref{fig:m1m2S} (a) shows the obtained bandstructure 
in the $k_z=0$ plane.
Its FS is shown in Fig. \ref{fig:fig6} (a).
The $d_{XZ}$-orbital ``pure-type'' band
corresponds to the present three-orbital model.
Its vHS energy is located at $E_{\rm vHS}^{XZ}\approx-0.1$.
Also, the $d_{YZ}$-orbital forms the ``mix-type'' band,
whose vHS energy is $E_{\rm vHS}^{YZ}\approx+0.1$.
In addition, the $d_{X^2-Y^2}$+$d_{3Z^2-R^2}$-orbital forms a 
pure-type band with the vHS energy $E_{\rm vHS}'\approx-0.05$.
Around M point, the $d_{XZ}$-orbital band near M point
is almost $k_z$-independent,
while other orbital bands exhibits small $k_z$-independences
(about $0.1\sim0.2$) in the band calculation.
}

\textcolor{black}{
Figure \ref{fig:m1m2S} (b) shows the bandstructure 
with introducing the $p$-orbital shift $\Delta E_{p}=-0.2$ eV.
Its FS is shown in Fig. \ref{fig:fig6} (b).
Here, the $d_{YZ}$-orbital (mix-type) band approaches $E_{\rm F}$
along the M-M' line,
consistently with the APRES measurement in Ref.
\cite{ARPES-VHSS}.
}

\textcolor{black}{
Here,
we derive the coefficients $m_1$ and $m_2$ defined as
${\bar M}_{\rm orb}= m_1 {\tilde{\bm \phi}}\cdot{\tilde{\bm \eta}} + m_2 {\tilde{\eta}_1}{\tilde{\eta}_2}{\tilde{\eta}_3}$,
where $\tilde{\eta_i}$ and $\tilde{\phi_i}$ are the 
order parameter of a specific $d$-orbital projected on the conduction band. 
The results for $d_{XZ}$-orbital are shown in 
Figs. \ref{fig:fig6} (e)-(h) in the main text.
(The weight of $d_\Gamma$-orbital ($\Gamma=XZ,YZ$, etc) on the 
conduction band is $W_{XZ}\approx0.7$, $W_{YZ}\approx0.3$, 
$W_{X^2-Y^2}\approx W_{3Z^2-R^2}\approx0.4$.)
We also calculate  $m_1$ and $m_2$ when the order parameters 
emerge on the $d_{YZ}$, $d_{X^2-Y^2}$, and $d_{3Z^2-R^2}$ orbitals.
The obtained results are shown in 
Figs. \ref{fig:m1m2S} (a)  ($\Delta E_{p}=0$) 
and (b) ($\Delta E_{p}=-0.2$),
as function of the electron filling $n$.
($n=31$ corresponds to undoped CsV$_3$Sb$_5$.)
It is found that large $m_1$ and $m_2$ are obtained 
when the current and bond orders emerge on various $d$-orbitals.
Thus, the present study is valid for various types of 
current order mechanisms, not restricted to 
Ref. \cite{Tazai-kagome2S}.
}

\subsection*{I: 
Increment of $T_{\rm c}$ under uniaxial strain in $3Q$ BO phase}\

\textcolor{black}{
In Ref. \cite{Moll-hzS}, it was proposed that
CsV$_3$Sb$_5$ is located at the quantum critical point of 
the current order ($T_{\rm c}^0\approx0$ in the absence of the uniaxial strain.
The field-induced ($h_z\sim9$T) current order at $T_{\rm c}\sim20$K
would be explained by the present theory.
In addition, Ref. \cite{Moll-hzS} reports the 
strain induced increment of $T_{\rm c}$
in the absence of the magnetic field.
It was explained in Ref. \cite{Moll-hzS}
based on the GL free energy analysis.
Under the uniaxial strain, the degeneracy of 
the current order transition temperature at $\q=\q_m$ ($m=1\sim3$),
$T_{\rm c}^{(m)}$, is lifted.
Then, $T_{\rm c}= \max_{m} T_{\rm c}^{(m)}$
will be larger than the original $T_{\rm c}^0$.
}

\textcolor{black}{
Here, we present that the strain induced change in the 4th order GL terms causes additional significant contribution to the increment of $T_{\rm c}$.
That is, the BO-induced suppression of the current order is drastically reduced by the strain.
Considering the $D_{6h}$ symmetry, 
we assume that the $E_{1g}$ symmetry strain $(\e,\e')$ 
induces the shift of the vHS energy levels as
$\Delta{\bm E}\equiv (\Delta E_A,\Delta E_B, \Delta E_C)= \a(\e,0,-\e)$ and
$\Delta{\bm E}'\equiv (\Delta E_A',\Delta E_B', \Delta E_C')=(\a'/2)(\e',-2\e',\e')$.
Then, the 4th order GL terms are given as
\begin{eqnarray}
\!\!\!\!\!\!\!\!\!\!\!\!\!\!\!\!\!\!\!\!
&&\sum_m^{1,2,3}\{ d_1^{(m)}\phi_m^4+d_3^{(m)}\eta_m^4
+2d_5^{(m)}\phi_m^2\eta_m^2\}
\nonumber \\
&&+\sum_m^{1,2,3}\{ d_2^{(o)}\phi_m^2\phi_n^2+d_4^{(o)}\eta_m^2\eta_n^2
+d_6^{(o)}(\phi_m^2\eta_n^2+\phi_n^2\eta_m^2) \},
\nonumber \\
\end{eqnarray}
where $n=m+1$ and $o=m-1$ mod 3.
When $\e=\e'=0$, $d_l^{(m)}=d_l$ for any $m$ and $l$.
Below, we show that
each $d_l^{(m)}$ has large $\e$- and $\e'$-linear terms
when $n\sim n_{\rm vHS}$.
}

\begin{figure}[htb]
\includegraphics[width=.99\linewidth]{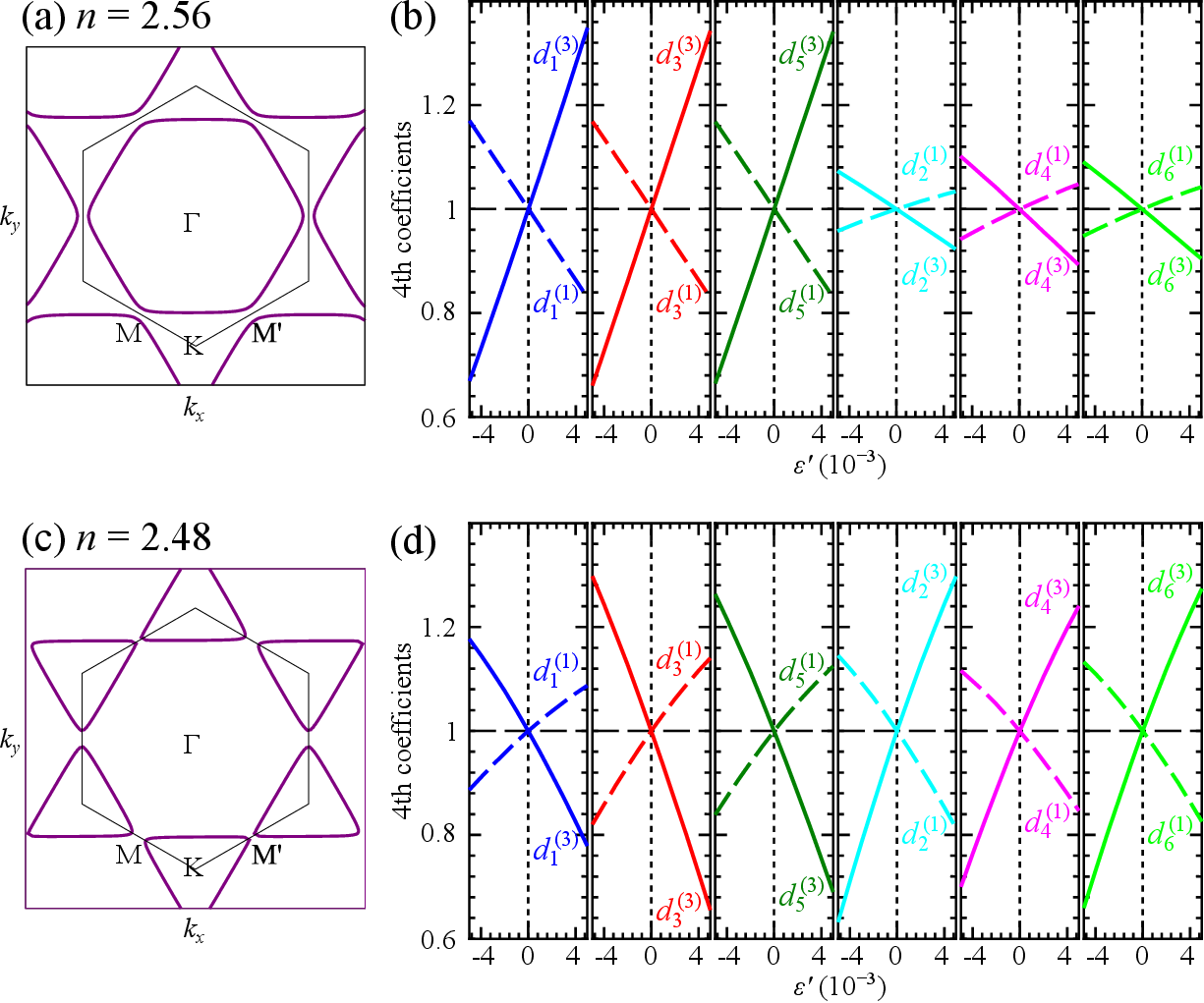}
\caption{
(a) FS for $n=2.56$ with $\e'=0.005$ and
(b) normalized $d_l^{(m)}$ ($m=1-3$; $l=1-6$)
as functions of the strain $\e'$
for $\Delta{\bm E}'= \a'(\e'/2,-\e',\e'/2$) with $\a'=1$.
Their $\e'$-linear terms give $g_l'$ in Eq. [\ref{eqn:DFe2}].
At $\e=0$, $(d_1,d_2,d_3,d_4,d_5,d_6) = (26.09,62.52,11.09,33.97,14.09,42.23)$.
(c) FS for $n=2.48$ with $\e'=0.005$ and 
(d) obtained $d_l^{(m)}$ as functions of the strain $\e'$.
At $\e'=0$, $(d_1,d_2,d_3,d_4,d_5,d_6) = (52.40,19.41,15.09,16.60,21.02,16.87)$.
In both (b) and (d),
the relations $d_l^{(1)}=d_l^{(2)}$ ($l=1-6$) and
$\d_{\e'} d_{l}^{(3)} = -2\d_{\e'} d_{l}^{(1)}$ holds.
The relation $\d_{\e'} d_{5}^{(m)} \sim -\d_{\e'} d_{6}^{(m)}$ 
means that $g_5'$ and $g_6'$ have the same sign.
Interestingly, $\d_{\e'} d_l^{(m)}$ for $n>n_{\rm vHS}$ in (b)
and that for $n<n_{\rm vHS}$ in (d) have opposite signs.
The used parameters are $T=0.01$ and $v=0.6$.
}
\label{fig:d-eS}
\end{figure}

\textcolor{black}{
Here, we calculate the coefficients $d_l^{(m)}$
in the three-orbital model
based on the Green function methods explained in the SI F.  
The used model parameters are $T=0.01$, $g=0.6$, $t'=0$, 
and $\a=\a'=1$.
We first study the case of $n=2.56 \ (>n_{\rm vHS})$,
whose FS at $\e'=0.005$ is shown in Fig. \ref{fig:d-eS} (a).
Figure \ref{fig:d-eS} (b) shows the obtained $d_l^{(m)}$ 
normalized by its $\e'=0$ value,
as functions of $\e'$ for $|\e'|\le0.005$.
We also study the case of $n=2.48\ (<n_{\rm vHS})$,
whose FS at $\e'=0.005$ is shown in Fig. \ref{fig:d-eS} (c).
Figure \ref{fig:d-eS} (d) shows the normalized $d_l^{(m)}$
as functions of $\e'$.
In both cases, the coefficients exhibits sizable $\e'$-linear terms,
by reflecting the large p-h asymmetry of the kagome lattice model.
}

\textcolor{black}{
In contrast, the irreducible susceptibility with the current form factor
$\chi_{lm,lm}^0(\q)=-T\sum_{\k,n}f_\q^{lm}(k)^*G_{ll}(\k+\q,\e_n)G_{mm}(\k,\e_n)f_\q^{lm}(k)$ 
exhibit much smaller relative change in the same range of $\e'$;
$\sim 3$\% for $n=2.56$ and $\sim5$\% for $n=2.48$.
The form factor ${\hat f}_\q(k)$ is introduced in the SI F.
(The change in $\chi_{lm,lm}^0(\q_i)$ modifies
the second order coefficient $a_{\rm c}$.)
The change in the 3rd order term is very small,
which is $O(\e^2)$ when $f_{ij}^{(m)}=\pm ig_{ij}^{(m)}$.
}

\textcolor{black}{
Now, we consider the strain-induced change 
in the 4th order free energy term.
By considering the symmetry property of the Feynman diagrams,
$\Delta F$ due to $\Delta{\bm E}\propto \e$ is given as
\begin{eqnarray}
\!\!\!\!\!\!\!\!\!\!\!\!\!\!\!\!\!\!\!\!
&&\Delta F(\e)=g_1(\phi_1^4-\phi_2^4)+g_3(\eta_1^4-\eta_2^4)
\nonumber \\
&&\ +2g_5(\phi_1^2\eta_1^2-\phi_1^2\eta_2^2)
\nonumber \\
&&\ +g_2(\phi_3^2\phi_1^2-\phi_2^2\phi_3^2)
+g_4(\eta_3^4\eta_1^2-\eta_2^2\eta_3^2)
\nonumber \\
&&\ +g_6(\eta_3^4\phi_1^2+\eta_1^4\phi_3^2-\eta_2^2\phi_3^2-\eta_3^2\phi_2^2) ,
\label{eqn:DFe}
\end{eqnarray}
where $g_l$ are $\e$-linear.
$g_l$ and $d_l^{(m)}$ are related as
$\e\cdot ( \d_\e d_l^{(1)},\d_\e d_l^{(2)},\d_\e d_l^{(3)})=g_l(1,-1,0)$
for $l$=odd, 
and $=-g_l(1,-1,0)$ for $l$=even.
Also, $\Delta F'$ due to $\Delta{\bm E}'\propto \e'$ 
is given as
\begin{eqnarray}
\!\!\!\!\!\!\!\!\!\!\!\!\!\!\!\!\!\!
&&\Delta F'(\e')=g_1'(\phi_1^4+\phi_2^4-2\phi_3^2)
+g_3'(\eta_1^4+\eta_2^4-2\eta_3^4)
\nonumber \\
&&\ +2g_5'(\phi_1^2\eta_1^2+\phi_2^2\eta_2^2-2\phi_3^2\eta_3^2)
\nonumber \\
&&\ +g_2'(\phi_2^2\phi_3^2+\phi_3^2\phi_1^2-2\phi_1^2\phi_2^2)
\nonumber \\
&&\ +g_4'(\eta_2^2\eta_3^2+\eta_3^2\eta_1^2-2\eta_1^2\eta_2^2)
\nonumber \\
&&\ +g_6'(\phi_2^2\eta_3^2+\phi_3^2\eta_2^2+\phi_3^2\eta_1^2+\phi_1^2\eta_3^2
-2\phi_1^2\eta_2^2-2\phi_2^2\eta_1^2) ,
\nonumber \\
\label{eqn:DFe2}
\end{eqnarray}
where $g_l'$ are $\e'$-linear.
$g_l'$ and $d_l^{(m)}$ are related as
$\e'\cdot ( \d_{\e'} d_l^{(1)},\d_{\e'} d_l^{(2)},\d_{\e'} d_l^{(3)})=g_l'(-1,-1,2)$ for $l$=odd, 
and $=-g_l'(-1,-1,2)$ for $l$=even.
}

\textcolor{black}{
Here, we consider the possible $3Q$ BO-current coexisting state
motivated by the experiment in Ref. \cite{Moll-hzS}.
As explained in Ref. \cite{Tazai-kagome2S},
in the case of $|{\bm \phi}|\gg|{\bm \eta}|$ at $\e=\e'=0$,
the 3Q BO ${\bm \phi}^0=(\phi^0,\phi^0,\phi^0)/\sqrt{3}$
coexists with the current order ${\bm \eta}\propto(\eta_1,\eta_2,-\eta_1-\eta_2)$
due to the energy gain by the 3rd GL terms with $b_1b_2<0$.
(The relation $b_1b_2<0$ is general \cite{Tazai-kagome2S}.)
Then, the BO-current coexisting state is nematic.
Here, we set ${\bm \eta}=(\eta,0,-\eta)/\sqrt{2}$
(or ${\bm \eta}=(\eta,-2\eta,\eta)/\sqrt{6}$)
without loss of generality.
}

\textcolor{black}{
Hereafter, we calculate the change in the current order transition 
temperature $\Delta {\bar T}_{\rm c}(\e) \ (\propto\e)$ under the $3Q$ BO phase.
(Below, we set $b_1=-b_2$ for simplicity to obtain the analytic expression.)
When ${\bm \eta}={\bm0}$,
we obtain the square of the $i$-th component of ${\bm \phi}$ as
$(\phi_i)^2=(\phi^0)^2/3+u_i\psi^2$ ($u_1=u_2=-1$ and $u_3=2$),
where $\psi^2=[(2g_1+g_2)/(2d_1-d_2)](\phi^0)^2 \ (\propto \e)$.
We consider $|{\bm\eta}|^2\sim O(\e)$ by assuming $T\sim T_{\rm c}$.
Then, the change in the free energy by $\Delta{\bm E}$ 
of order $O(\eta^2)$ is obtained as
$\Delta F=-[(2g_5+g_6)+D(2g_1+g_2)]((\phi^0)^2/6)\cdot\eta^2$,
where $D\equiv (2d_5-d_6)/(2d_1-d_2)$.
In the present numerical study, $D\sim1$ and $(2g_5+g_6)\approx(2g_1+g_2)$.
(Exactly speaking, $D=1.36$ (0.29) for $n=2.56$ (2.48).)
}

\textcolor{black}{
In the BO phase at $\e=0$,
the original 2nd order GL coefficient $a_{\rm c}$ 
is changed by $\phi^0\ne0$ as
\begin{eqnarray}
{\bar a}_{\rm c}(0)&=& a_{\rm c} + 2(d^5+d^6)((\phi^0)^2/3)
\nonumber \\
&\propto& T-{\bar T}_{\rm c}(0),
\label{eqn:ac0S}
\end{eqnarray}
where $|\phi^0|\gtrsim T_{\rm b}$ when $T\ll T_{\rm b}$.
For finite $\e$, it is changed by $\Delta F(\e)$ as
\begin{eqnarray}
{\bar a}_{\rm c}(\e)&=& {\bar a}_{\rm c}(0)-[(2g_5+g_6)+D(2g_1+g_2)]((\phi^0)^2/6) 
\nonumber \\
&\propto& T-{\bar T}_{\rm c}(\e),
\label{eqn:acS}
\end{eqnarray}
The current order appears in the BO phase
when Eq. [\ref{eqn:acS}] becomes negative.
Therefore, ${\bar T}_{\rm c}$ will increase in proportion to $\e$ if $(2g_5+g_6)>0$.
(If $(2g_5+g_6)<0$, $T_{\rm c}$ increases
when ${\bm\eta}\propto(0,\eta,-\eta)$.)
}

\textcolor{black}{
In the same way, ${\bar a}_{\rm c}(\e)$ is modified by $\Delta F'(\e')$ as
\begin{eqnarray}
{\bar a}_{\rm c}(0)-[(2g_5'+g_6')+D(2g_1'+g_2')]((\phi^0)^2/6).
\label{eqn:acS}
\end{eqnarray}
Note that $g_5'=-\e'\d_{\e'}d_5^{(1)}$ and 
$g_6'=\e'\d_{\e'}d_6^{(1)}$ has the same sign 
according to Fig. \ref{fig:d-eS},
which is naturally expected analytically.
}

\textcolor{black}{
Considering the drastic $\e'$-dependence of $d_l^{(m)}$
obtained in Fig. \ref{fig:d-eS} (a),
in collaboration with the change in $a_{\rm c}$ 
already discussed in Ref. \cite{Moll-hzS},
sizable strain-induced increment of 
the current order transition temperature $T_{\rm c}$ 
reported in Ref. \cite{Moll-hzS}
will be realized in the present mechanism.
}

\textcolor{black}{
In the case of $f_{ij}^{(m)}=\pm ig_{ij}^{(m)}$,
$g_{l}$ and $g_{l}'$ coincide with $l$=even and $l$=odd, respectively.
Then, $D=1$ and $(2g_5+g_6)=(2g_1+g_2)$.
The main results are valid even in this case.
}


\end{document}